\documentclass[fleqn,usenatbib]{mnras}

\usepackage{newtxtext,newtxmath}

\usepackage[T1]{fontenc}

\DeclareRobustCommand{\VAN}[3]{#2}
\let\VANthebibliography\thebibliography
\def\thebibliography{\DeclareRobustCommand{\VAN}[3]{##3}\VANthebibliography}

\usepackage{graphicx}	
\usepackage{amsmath}	
\usepackage{epstopdf}
\usepackage{color}
\usepackage{subcaption}
\usepackage{booktabs}
\usepackage{orcidlink}

\newcommand{\xmm}{{\it XMM-Newton}}
\newcommand{\swift}{{\it Swift}}
\newcommand{\nicer}{{\it NICER}}
\newcommand{\nustar}{{\it NuSTAR}}
\newcommand{\swiftbat}{{\it Swift}/BAT}

\newcommand{\swiftuvot}{{\it Swift}/UVOT}
\newcommand{\maxi}{{\it MAXI}/GSC}
\newcommand{\chandra}{{\it Chandra}}

\newcommand{\isis}{{ISIS}}

\newcommand{\source}{MAXI J1820+070}

\title[The origin of optical emission lines in XRBs]{The origin of optical emission lines in the soft state of X-ray binary outbursts: the case of MAXI J1820$+$070}

   \author[K.~I.~I.~Koljonen et al.]{K.~I.~I.~Koljonen$^{\orcidlink{0000-0002-9677-1533}}$,$^{1,2,3}$\thanks{E-mail: karri.koljonen@ntnu.no}
   K.~S.~Long$^{\orcidlink{}}$,$^{4,5}$\thanks{E-mail: long@stsci.edu}
   J.~H.~Matthews$^{\orcidlink{0000-0002-3493-7737}}$,$^{6}$
   and C.~Knigge$^{\orcidlink{0000-0002-1116-2553}}$$^{7}$
   \\
   $^{1}$Institutt for Fysikk, Norwegian University of Science and Technology, H{\o}gskloreringen 5, Trondheim, 7491, Norway\\
   $^{2}$Finnish Centre for Astronomy with ESO (FINCA), University of Turku, V\"ais\"al\"antie 20, 21500 Piikki\"o, Finland \\
   $^{3}$Aalto University Mets\"ahovi Radio Observatory, PO Box 13000, FI-00076 Aalto, Finland \\
   $^{4}$Space Telescope Science Institute, 3700 San Martin Drive, Baltimore, MD 21218, USA \\
   $^{5}$Eureka Scientific, Inc.,2452 Delmer Street, Suite 100, Oakland, CA 94602-3017, USA \\
   $^{6}$Department of Physics, Astrophysics, University of Oxford, Denys Wilkinson Building, Keble Road, Oxford, OX1 3RH, UK \\
   $^{7}$School of Physics and Astronomy, University of Southampton, Highfield, Southampton, SO17 1BJ, UK
}

\date{\today
}

\pubyear{2023}

\begin{document}
\label{firstpage}
\pagerange{\pageref{firstpage}--\pageref{lastpage}}
\maketitle
\begin{abstract}  
The optical emission line spectra of X-ray binaries (XRBs) are thought to be produced in an irradiated atmosphere,  possibly the base of a wind, located above the outer accretion disc. However, the physical nature of -- and physical conditions in -- the line-forming region remain poorly understood. Here, we test the idea that the optical spectrum is formed in the transition region between the cool, geometrically thin part of the disc near the mid-plane and a hot, vertically extended atmosphere or outflow produced by X-ray irradiation. We first present a VLT X-Shooter spectrum of XRB MAXI J1820+070 in the soft state associated with its 2018 outburst, which displays a rich set of double-peaked hydrogen and helium recombination lines. Aided by ancillary X-ray spectra and reddening estimates, we then model this spectrum with the Monte Carlo radiative transfer code {\sc Python}, using a simple biconical disc wind model inspired by radiation-hydrodynamic simulations of irradiation-driven outflows from XRB discs. Such a model can qualitatively reproduce the observed features; nearly all of the optical emission arising from the transonic `transition region' near the base of the wind. In this region, characteristic electron densities are on the order of 10$^{12-13}$ cm$^{-3}$ , in line with the observed flat Balmer decrement (H$\alpha$/H$\beta \approx 1.3$). We conclude that strong irradiation can naturally give rise to both the optical line-forming layer in XRB discs and an overlying outflow/atmosphere that produces X-ray absorption lines.
\end{abstract}

\begin{keywords}
Accretion, accretion discs -- binaries: close -- stars: individual: MAXI J1820+070 -- stars: winds, outflows -- X-rays: binaries
\end{keywords}



\section{Introduction} \label{introduction}
X-ray binaries (XRBs) are systems where a compact object accretes material from a non-degenerate companion star. The accreting gas exhibits a large range of temperatures and densities -- primarily depending on the distance to the compact object -- allowing us to study a wide variety of astrophysical plasmas in a single source. XRBs also drive outflows from their discs in the form of fast, strongly collimated jets and slower, less collimated disc winds \citep[e.g.][]{fender04,ponti12}. The disc winds are observationally more elusive than jets, but their mass-loss rate is larger and may occasionally even exceed the accretion rate. As a result, disc winds have the potential to severely affect the overall accretion flow \citep{munozdarias16,casares19}. They might also extract angular momentum from the disc regulating the evolution of XRB outbursts \citep{tetarenko18}.     

Evidence for equatorial disc winds has mostly been accumulated via X-ray spectroscopy obtained during outbursts in the form of blueshifted absorption lines associated with highly ionized absorbers \citep[\ion{Fe}{xxv} K$\alpha$ / \ion{Fe}{xxvi} Ly$\alpha$;][and references therein]{ponti16,diaztrigo16}. Interestingly, these absorption lines are typically observed in the so-called soft X-ray state, during which the luminosity is dominated by thermal emission from the disc. However, they are rarely present during the so-called hard X-ray state, when non-thermal emission from a hot, optically thin plasma dominates the X-ray spectrum and radio jets are observed \citep{miller06,neilsen09,neilsen11,ponti12,ponti14}. This has sparked discussion of whether the jets and winds exclude each other \citep{neilsen09}, or whether they can coexist as seems to be the case at least in a few sources \citep{rahoui14,munozdarias16,homan16,munozdarias17,allen18}. 

In addition to X-rays, evidence of disc winds has also been seen in optical spectra, most convincingly in the form of P-Cygni-like, blueshifted broad absorption profiles associated with hydrogen and helium optical lines 
\citep{munozdarias16,munozdarias18,munozdarias19,charles19,jimenezibarra19}. Recently, similar disc wind signatures have also been detected in the profiles of {\em ultraviolet} resonance lines \citep{castro22}.
However, contrary to X-ray observations, these lower-energy wind signatures are seen preferentially during the hard state. This raises an interesting question of whether the disc wind is present throughout the outburst, but its properties change between the soft and hard states. Such changes could be associated with the disc geometry \citep{ueda10,miller12,ponti12} and/or the ionisation state of the wind \citep{jimenezgarate01,ueda10,diaztrigo12,chakravorty13, diaztrigo14,higginbottom15,diaztrigo16,bianchi17}.

\source\/ (ASASSN-18ey) was first detected at the start of its outburst in early 2018 \citep{tucker18} and became one of the brightest XRB outbursts ever observed, with an X-ray flux reaching $\sim$4 Crab and an optical magnitude reaching 11.2 \citep{shidatsu19}. However, as the source is relatively close by, 3.0$\pm$0.3 kpc \citep{atri20}, the inferred peak luminosity was quite typical for an XRB outburst, at about 15\% of the Eddington luminosity. Since disc winds are preferentially detected in systems viewed more edge-on \citep{ponti12, higginbottom20}, the relatively high inclination inferred for \source\ -- 
 $i>$60 deg \citep{atri20,torres20} -- makes it a promising target for observational studies aimed at these outflows. Indeed, clear accretion disc wind features (P-Cygni profiles and broad emission line wings) were observed in the hard state during both outburst rise and decay \citep{munozdarias19}. In addition, wind features were found during the soft state in the infrared Pa$\gamma$ and Pa$\beta$ lines, indicating that the outflow was present throughout the outburst \citep{sanchezsierras20}. Clear absorption lines in the X-ray spectra have not been found (partly because the outburst was too bright to be observed with \xmm\/ and \chandra), but some \nicer\/ soft state spectra show dips at $\sim$7 keV that can be fitted with an outflowing ionized absorber \citep{fabian20}. 

Most of the optical studies on XRB accretion disc winds have concentrated on the effects of the wind on single lines in the form of P-Cygni profiles and broad emission line wings. By contrast, the blue continuum and double-peaked emission lines are usually taken to arise directly from the irradiated accretion disc. However, it is worth asking if the wind could affect the observed spectrum more generally, imprinting it with emission lines and an additional continuum component. This is especially important given the evidence that the disc wind is present at all times. Models like this have been explored in the context of accreting white dwarfs with promising results \citep[e.g.][]{matthews15}. However, in XRBs, the optical continuum is typically described simply with a power law model, while the lines are modeled individually.    

In this paper, we study the possible effect of a disc wind -- or, more generally, an extended disc atmosphere -- on the UV, optical, and infrared (UVOIR) spectrum of \source\ during the soft state of its outburst. The underlying irradiating continuum is estimated by modelling quasi-simultaneous multi-wavelength observations obtained by ESO's Very Large Telescope (VLT), \swift, \nicer, and \nustar. We then use \textsc{python}, a Monte Carlo ionisation and radiative transfer code to reproduce the observed line spectrum and to derive the physical parameters of the disc wind/atmosphere.

\section{Observations} \label{observations}
 
\begin{figure}
 \centering
 \includegraphics[width=\linewidth]{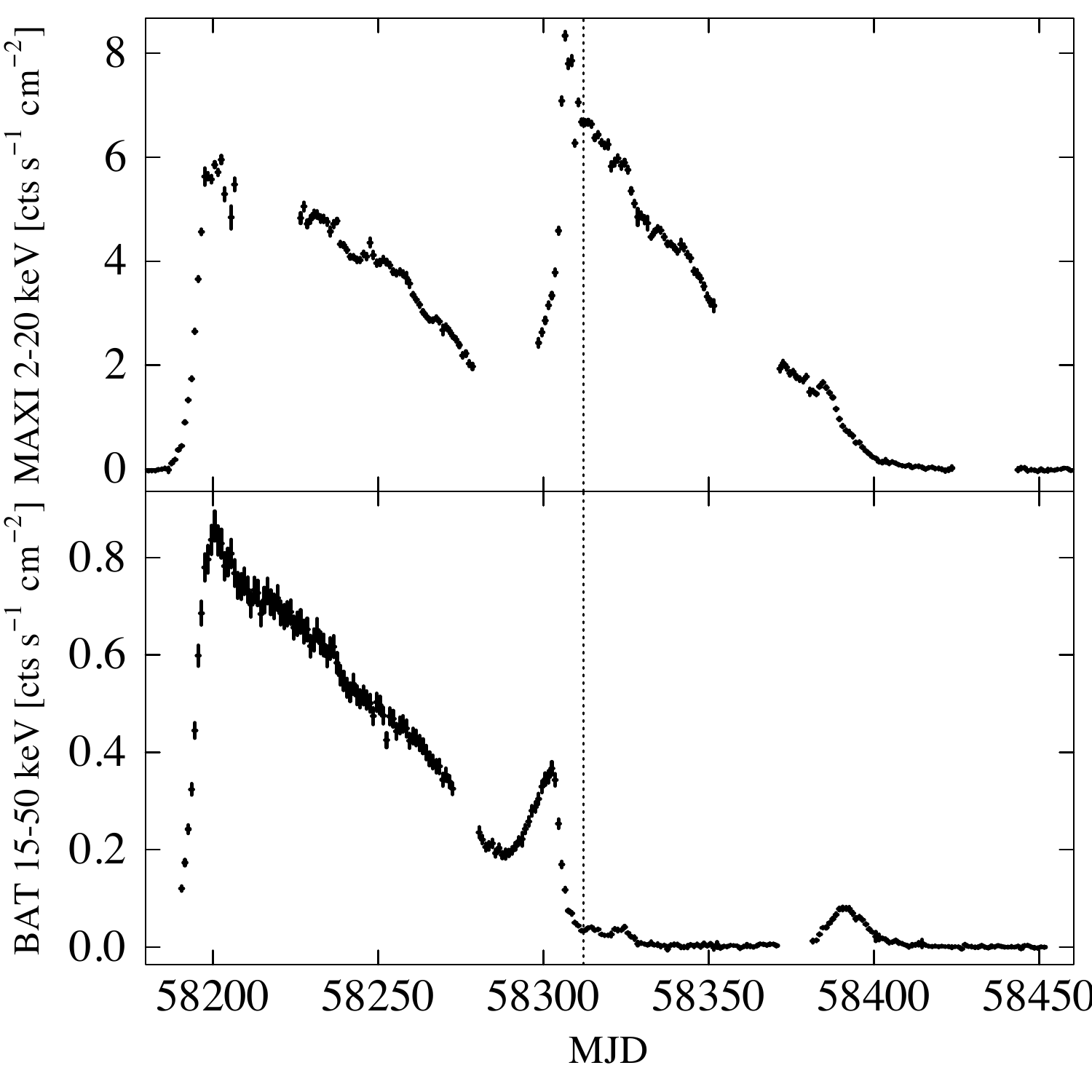}
 \caption{\maxi\/ and \swiftbat\/ lightcurves of \source. The time of the X-Shooter observation taken after the change from the hard state to the soft state is marked with a dotted line.}
 \label{bat}
\end{figure}

 We carried out a multi-epoch observing campaign with VLT/X-Shooter covering the 2018 outburst of \source. In this paper, we consider only one epoch of this campaign, which coincided with the soft state. During this epoch, the broadband spectrum was dominated mainly by a thermal disc component (Fig. \ref{bat}). 
 
 In addition, we analyzed UV and X-ray pointed observations taken at approximately the same time as the X-Shooter data, allowing us to construct a full spectral energy distribution (SED). We downloaded \swift, \nicer, and \nustar\/ pointing observations from the High Energy Astrophysics Science Archive Research Center (HEASARC) that were as close as possible to the time of the X-Shooter observation.

\subsection{X-Shooter}

The X-Shooter observation considered here was obtained during the soft state starting on MJD 58312.118 (2018-07-13 02:50). The observation consisted of four exposures of $\sim$10 minutes each, arranged in AB pairs alternating between source and sky positions. We used a nod throw length of five arcseconds and a jitter box of one arcsecond. The total observing time, including overheads, was $\sim$45 minutes. 

The slit acquisition was carried out with the i'-band filter, which we also used to correct the flux level of the X-Shooter spectra due to slit losses. The `i'-band magnitude of \source\/ was 13.0 without correcting for the interstellar reddening. The slit widths used were 1.3, 1.3, and 1.2 arcseconds for UV, VIS, and NIR arms, respectively. The readout mode of the detector was 100k/1pt/hg/1x2. We reduced the X-Shooter data with the ESO pipeline v3.5.0 in EsoReflex. The telluric absorption was corrected using \textsc{molecfit} \citep{smette15}. 

To facilitate the broadband continuum fitting presented in Section \ref{sec:continuum}, we formed a line/edge-free continuum spectrum by selecting first only continuum regions from the X-Shooter spectrum and interpolating over the line/edge regions. We  binned the spectrum to 100 logarithmically spaced wavelength bins and assumed a 5\% error for each data point.   

\subsection{\swiftuvot}

\begin{table}
\centering
\caption{\swiftuvot\/ photometry of \source\/ on 2018-07-13. The magnitudes and fluxes correspond to reddened values.} \label{uvot}
 \begin{tabular}{llllll}
 \hline\hline
 Filter & Freq. & Exp. & Mag. & Flux & Err. \\
 & [10$^{14}$ Hz] & [s] & [mJy] & [mJy] & [mJy] \\
\hline
 V & 5.55 & 135 & 12.99 & 23.0 & 0.7  \\
 U & 8.56 & 135 & 13.12 & 20.8 & 0.8  \\
 UW1 & 11.6 & 270 & 13.48 & 14.5 & 0.7 \\
 UM2 & 13.5 & 229 & 13.76 & 11.4 & 0.4 \\
 UW2 & 14.8 & 540 & 13.65 & 12.5 & 0.6 \\
 \hline
 \end{tabular}
\end{table}

There were two \swift\/ pointed observations of \source\/ on the date of the X-Shooter observation: pointing 00010754003 with observation start time MJD 58312.208 (2018-07-13 05:00) and pointing 00010754004 with observation start time MJD 58312.341 (2018-07-13 08:11). While the closest \swift\/ observation to the X-Shooter data was pointing 00010754003, it contained only one UVOT exposure (taken in the U-band). We, therefore, chose to use the second pointing, obtained three hours later, which contained exposures in five UVOT bands: V, U, UW1, UM1, and UW2. The U-band magnitudes were consistent at both pointings.

The magnitudes and flux densities of \source\/ were calculated using the \textsc{uvotsource} task in \textsc{heasoft 6.26.1} for each filter. For extracting the source counts, we used a circular aperture with a radius of five arcseconds centered on the location of the source (RA 18:20:21.4, DEC +07:10:53.1). The background was estimated using a circular source-free aperture with a radius of 12 arcseconds located close to \source. Table \ref{uvot} shows the results from the photometry. 

\subsection{\nustar}

The \nustar\/ observation closest to the X-Shooter data in time was pointing number 0401309025, with an observation start time of MJD 58314.744 (2018-07-15 17:51). While this observation was taken $\sim$2.5 days later, the \maxi\/ and \swiftbat\/ count rates remained consistent across these dates, so we do not expect any significant change in the X-ray spectral shape or normalisation. Adding the \nustar\/ data to the broadband spectral energy distribution provides essential constraints on the amount of hard X-ray emission and the irradiation of the accretion disc.

We reduced the \nustar\/ data from the two focal plane modules (FPMA and FPMB) using the \textsc{nupipeline} task contained in \textsc{heasoft 6.26.1}. We used a circular source region with a 160 arcsec radius centered on the location of the source. We estimated  the background using a circular aperture with a 160 arcsec radius located in a source-free region of the image. We ran the pipeline with the parameters \textsc{tentacle=`yes'} and \textsc{saamode=`optimized'}. The former requires a simultaneous increase in the CdZnTe detector event count rates and the observed shield single rates, while the latter allows identification and flagging of time intervals in which the CdZnTe detector event count rates show an increase when the spacecraft enters the South Atlantic Anomaly (SAA). 

We extracted averaged spectra from the two detectors with the task \textsc{nuproducts} contained in \textsc{heasoft 6.26.1}. The broadband (3--79 keV) \nustar\/ count rate ranged between 490--530 cts/s. For X-ray modeling, we binned the data to a minimum signal-to-noise ratio (S/N) of 30 across the full 3--79 keV band.  

\subsection{\nicer}

The \nicer\/ pointing closest in time to the \nustar\/ dataset was obsid 1200120207 and took place on MJD 58314.181 (2018-07-15 04:20). The on-source exposure time was 4.6 ksec. 

We reduced the observation using \textsc{nicerdas} version 10, using the \textsc{nicerl2} script with default parameters. We extracted the X-ray spectra and accompanying detector response files using the \textsc{nicerl3-spect} script and opted for SCORPEON (version 22) background modeling. For the spectral analysis, we binned the data to S/N = 30. We performed spectral fitting using the Interactive Spectral Interpretation System (\isis; Houck 2002) and estimated the errors on the best-fit parameter values via Monte Carlo analysis. In the modeling, we added a constant factor to account for the flux difference between the X-ray detectors.

\begin{figure*}
 \centering
 \includegraphics[width=0.95\linewidth]{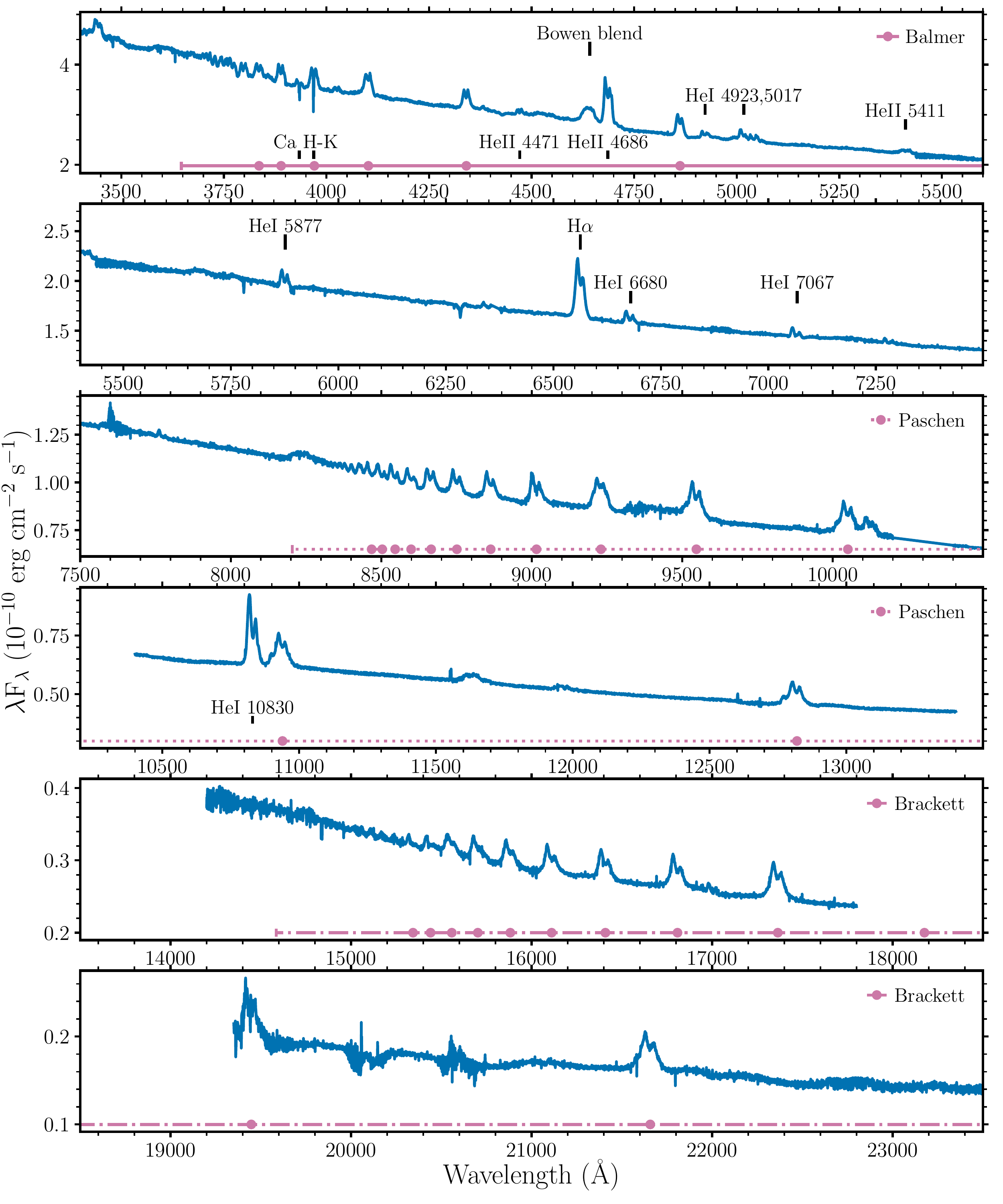}
 \caption{The de-reddened optical-to-near-infrared spectrum obtained with X-Shooter shows a rich emission line spectrum and blue continuum. Prominent emission lines are marked, and the Balmer, Paschen, and Brackett series of Hydrogen emission lines are labelled with pink dots. 
 }
 \label{xs}
\end{figure*} 

\section{Results} \label{results}

\subsection{Line spectrum}

The X-Shooter spectrum is rich in optical and infrared emission lines, as shown in Fig. \ref{xs}. The underlying continuum rises towards the UV, as expected for emission produced by an accretion disc. The spectrum shows a remarkable collection of emission features, most of which correspond to recombination lines associated with hydrogen and helium. For example, the Balmer, Paschen, and Brackett series of hydrogen are all identifiable, down to transitions 10--2, 18--3, and 18--4, respectively. The corresponding photoionisation edges are also visible and are observed in emission. This immediately implies that there is a significant recombination continuum contribution to the broadband optical SED. Evidently,  the material responsible for generating the emission lines  also affects the overall continuum shape.

All of the hydrogen lines are double-peaked, except Pa$\beta$ and Pa$\gamma$, which display a weaker third peak in the blue line wing. The line profile of all hydrogen lines is relatively uniform, with the blue peak being more substantial. The only exceptions are the higher-order Balmer transitions, which show peaks of roughly equal strength. 

In addition to the hydrogen lines, several helium emission lines are also visible, notably \ion{He}{i} $\lambda$10830 and \ion{He}{ii} $\lambda$4686. Weaker helium lines can be found at \ion{He}{i} $\lambda$4026, \ion{He}{i} $\lambda$4388, \ion{He}{i} $\lambda$4471, \ion{He}{i} $\lambda$4922, \ion{He}{i} $\lambda$5016, \ion{He}{i} $\lambda$5048, \ion{He}{i} $\lambda$5876, \ion{He}{i} $\lambda$6678, \ion{He}{i} $\lambda$7065, \ion{He}{i} $\lambda$7281, \ion{He}{i} $\lambda$11969, \ion{He}{i} $\lambda$17002, \ion{He}{ii} $\lambda$5412, \ion{He}{ii} $\lambda$10124, and \ion{He}{ii} $\lambda$11626. The shapes of the helium lines mirror those of the hydrogen lines in displaying stronger blue peaks. 

Two prominent Bowen fluorescence features are also visible: \ion{O}{III} $\lambda$3444 and the \ion{O}{III}/\ion{N}{III} blend near $\lambda$4650. The formation of these features requires a strong extreme-UV source, so their presence is consistent with our observation in the soft state (c.f. Figure~3). In addition, the \ion{Ca}{ii} H\&K lines are observed both in emission and absorption (with the absorption likely being produced in the interstellar medium).

Two emission lines escaped our classification that are located at roughly $\lambda$6338 and $\lambda$7762. We also identify several diffuse interstellar band (DIB) lines in the X-Shooter spectra, centered at $\lambda$5780, $\lambda$5797, $\lambda$6202, $\lambda$6270 and $\lambda$6284.   

\subsection{Reddening} \label{redden}

It is important to adopt appropriate estimates of extinction ($A_{\rm V}$) and reddening ($E_{\rm B-V}$) when modelling optical data. As discussed in more detail below, three different methods suggest $E_{\rm B-V} \simeq 0.2$ for \source. We adopt the standard value of $R_{\rm V}$ = $A_{\rm V}/E_{\rm B-V}$ = 3.1 throughout our analysis. 

\subsubsection{HI column density}
\label{HI}

Extinction and reddening are strongly correlated with neutral atomic hydrogen column density ($N_H$) along any given line of sight \citep[e.g.][]{bohlin78,liszt14,friedman11}. In the case of \source, $N_H$ can be obtained from 21-cm radio observations by fitting the interstellar absorption of the X-ray spectrum and by measuring the equivalent widths of diffuse interstellar bands (DIBs).
\footnote{We use DIB EWs to estimate $N_H$, rather than $E_{\rm B-V}$, since the former correlation is much tighter \citep{friedman11}.}

First, \citet{HI4PI} provide an estimate of 1.3$\times$10$^{21}$ atoms cm$^{-2}$ in the direction of \source\, based on the high-resolution 21-cm radio observations. Second, in Section \ref{sec:continuum}, we fit the absorbed X-ray spectrum using an interstellar absorption model \textsc{tbnew} with elemental abundances from \citet{wilms00}. This analysis yields an estimate of N$_{\mathrm{H}}$ = 1.3 $\times$ 10$^{21}$ atoms cm$^{-2}$. Third, we have estimated the equivalent width of the $\lambda$5780 line in our X-Shooter data from a Voigt profile fit. This yielded EW($\lambda$5780)  = 0.188 \AA, which, based on the correlation between EW and NH established by \cite{friedman11}, suggests a value of N$_{\mathrm{H}}$ = 1.4$\pm$0.4 $\times$ 10$^{21}$ atoms cm$^{-2}$. Thus all three methods yield similar values of N$_{\mathrm{H}}$ .

The reddening can then be estimated from $N_H$ via the N$_{\mathrm{H}}$/E$_{\mathrm{B-V}}$ relation \citep{bohlin78,liszt14}. Assuming a conversion of N$_{\mathrm{H}}$/E$_{\mathrm{B-V}}$ = 8.3$\times \rm 10^{21} \:cm^{-2} \, mag^{-1}$ at |b| <~ 30\degr \citep{liszt14}, the implied E$_{\mathrm{B-V}}$ is 0.17. This is somewhat less than would be derived from the often quoted relationship of N$_{\mathrm{H}}$ = 5-6 $\times \rm 10^{21} \:cm^{-2} \, mag^{-1}$ \citep{bohlin78}, which would lead to E$_{\mathrm{B-V}}$ between 0.23 and 0.28. 

\subsubsection{Dust maps}

We can also estimate the reddening and extinction towards \source\ from models of the Galactic dust distribution. 
We used the GALExtin service \citep{amores21} for this purpose, which provides convenient access to several such models.
\footnote{It should be acknowledged here that these dust model estimates cannot be considered strictly independent since the models tend to be calibrated with the same core data sets. Moreover, some models -- e.g., the \citet{amores05} one -- are at least partially calibrated via HI observations. Such resulting estimates are therefore also not strictly independent from those obtained in Section~\ref{HI}. Nevertheless, it is at least somewhat reassuring that the dust model estimates are consistent both with each other and with the $N_H$-based estimates.}
For the distance and direction of \source\, the model of \citet{amores05} gives A$_{\mathrm{V}}=0.59$ mag. For our standard value of $R_{\mathrm{V}}= A_V/E_{\mathrm{B-V}} = 3.1$, this corresponds to E$_{\mathrm{B-V}}=0.19$. Similarly, the model of \citet{drimmel03} gives E$_{\mathrm{B-V}}=0.22$, \citet{schlafly14} and \citet{planck14} both suggest E$_{\mathrm{B-V}}=0.25$, and the model of \citet{green18} yields E$_{\mathrm{B-V}}=0.17$. Thus dust models also suggest a value of E$_{\mathrm{B-V}}\simeq 0.2$ for \source. 

\subsubsection{The 2200~\AA\/ feature}

A final reddening estimate can be obtained from the ultraviolet continuum shape. During its outburst, we observed \source\ twice with the {\em Hubble Space Telescope}, providing two epochs of far- and near-ultraviolet spectroscopy. A complete analysis of the resulting data will be presented separately (Georganti et al. 2023, in preparation), but the strength of the well-known $2200$~\AA\/ dust absorption feature in these observations also suggests E$_{\mathrm{B-V}}\simeq 0.2$.

\subsection{Broadband continuum} \label{sec:continuum}

\begin{table}
 \centering
 \caption{System parameters of MAXI J1820+070 found in the literature. 
 We adopt an inclination of $65^\circ$ and $M_{\rm BH} = 8.5~M_{\odot}$ in our modelling.
 } \label{system_param}
 \begin{tabular}{lcrc}
 \hline\hline
 Parameter & Unit & Value & Ref. \\
 \hline
 Distance & kpc & 2.96$\pm$0.33 & 1 \\
 Jet inclination & deg & 63$\pm$3 & 1 \\
 Orbital inclination & deg & 62 $<i<$ 81 & 2 \\ 
 $M_{\rm BH}$ (jet) & $M_\odot$ & 8.5$^{+0.8}_{-0.7}$ & 2 \\
 $M_{\rm BH}$ (orbit) & $M_\odot$ & 6.0 $<M_{\rm BH}<$ 8.1 & 2 \\
 $q$ & & 0.072$\pm$0.012 & 2 \\
 \hline  
 \multicolumn{4}{p{0.95\linewidth}}{\textbf{References:} 1) \citet{atri20}, 2) \citet{torres20}}
 \end{tabular}
\end{table} 

\begin{figure*}
 \centering
 \includegraphics[width=0.8\linewidth]{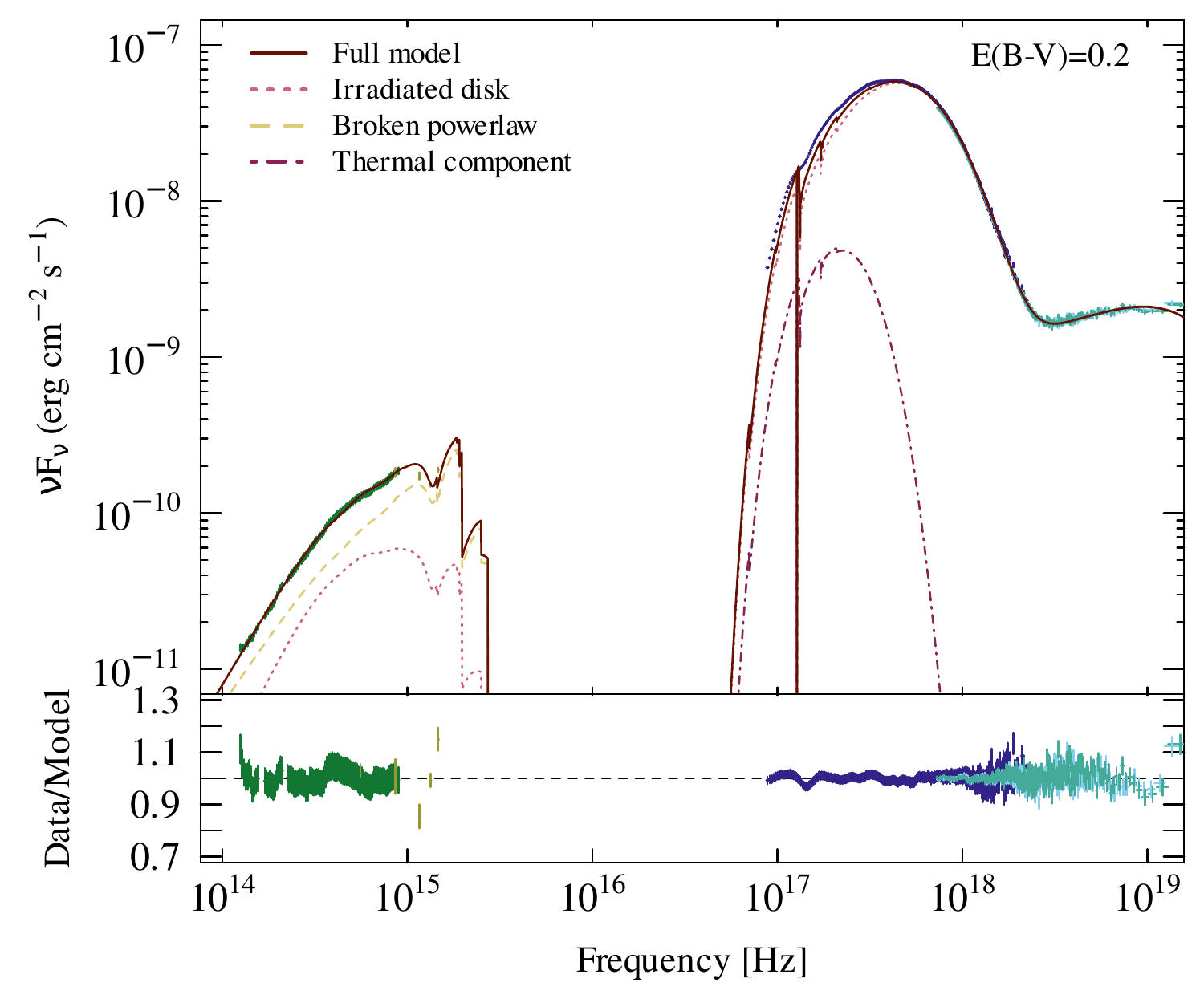}
 \caption{
 Broadband data fitted with the model described in Section \ref{sec:continuum}. Dark green: X-Shooter data, light green: \swiftuvot\/ data, dark blue: \nicer\/ data, light blue: \nustar\/ data, solid curve: full absorbed model as tabulated in Table \ref{params}, dotted curve: disc irradiation component, dashed curve: broken power-law in the UVOIR region (possibly a component arising from the wind/disc atmosphere), dot-dashed curve: `soft excess' (possibly a component arising from the wind/disc atmosphere).
 }
 \label{bbcont}
\end{figure*}

\begin{table*}
\centering
\caption{Full SED model parameters.} \label{params}
\begin{tabular}{llllll}
\multicolumn{6}{l}{\textsc{constant} $\times$ \textsc{redden} $\times$ \textsc{tbnew} $\times$ (\textsc{optxrplir}+\textsc{bknpower}+\textsc{bbody})} \\
\hline\hline
\textsc{tbnew} & $N_{\mathrm{H}}$ \\
& (10$^{20}$ cm$^{-2}$) \\ 
\cmidrule{2-2}
& 12.95$\pm$0.19  \\  
\hline\hline
\textsc{optxrplir} & $L$ & $kT_{\mathrm{bb}}$ & $r_{\mathrm{pl}}$ & $\Gamma_{\mathrm{pl}}$ \\ 
& ($L_{\mathrm{Edd}}$) & (keV) & ($R_{g}$) & \\
\cmidrule{2-5}
& 0.1266$\pm$0.0002 & 1.00$\pm$0.01 & 8.98$^{+0.05}_{-0.04}$ & 1.70$\pm$0.01 \\
\cmidrule{2-6}
& $r_{\mathrm{comp}}$ & $kT_{e}$ & $\tau$ & $f_{\rm out}$ & log $r_{\rm out}$ \\
& ($R_{g}$) & (keV) & & & ($R_{g}$) \\
\cmidrule{2-6}
& 17.2$\pm$0.5 & 20.5$\pm$2 & 7.8$\pm$0.1 & 0.013$\pm$0.004 & 5.29$\pm$0.06 \\
\hline\hline
\textsc{bknpower} & norm & $\beta$ \\
\cmidrule{2-3}
& 5170$^{+1989}_{-1244}$ & 0.19$^{+0.04}_{-0.06}$ \\
\hline\hline
\textsc{bbody} & norm & $kT_{\rm bb}$ \\
& & (keV) \\
\cmidrule{2-3}
& 0.130$\pm$0.007 & 0.187$\pm$0.004 \\ 
\hline\hline
\multicolumn{6}{l}{Fit quality: $\chi^{2}$/d.o.f. = 1147/1281; \, $\chi^{2}_{\mathrm{red}}$ = 0.90} \\ 
\multicolumn{6}{l}{Fixed parameters: $\theta_{\mathrm{inc}}=65$ deg; \, $M_{\rm BH}=8.5 M_{\odot}$; \, $d=3$ kpc; \, $a_{*}=0$; \, $a_{\rm out}=0.9$; \, $E_{\rm B-V}$=0.2} \\ 
\end{tabular}
\end{table*}

In the soft state of XRBs, the X-ray spectrum is dominated by a thermal component peaking around 1 keV. The thermal component is typically accompanied by a weak hard Compton tail extending to higher energies. These components are often interpreted as arising, respectively, from a thermal, multicolor accretion disc and a non-thermal, optically thin region possibly associated with reconnection sites above the thermal disc \citep{haardt94} or with a small `corona' close to the black hole.  In addition to the canonical disc and non-thermal component, an additional thermal component is needed to fit the soft state X-ray spectrum of \source. This has been suggested to arise from a plunge region close to the black hole \citep{fabian20,zhu12}. On the other hand, similar thermal components have been reported before from other XRBs \citep{zycki01,chiang10,shaw16}, and another possibility for it could be inner disc irradiation or enhanced dissipation in the upper layers of the disc (\citealt{davis05}; see also discussion in \citealt{done12}). 

During XRB outburst soft states, the strongest component in the UVOIR spectral region likely arises from the irradiated outer disc illuminated by the X-ray emission emanating from close to the black hole. The optical emission from the companion star is also much weaker, by several orders of magnitude \citep{torres20}. Typically, the viscous disc spectrum in the UVOIR is a power law with an index close to $\beta = 1/3$; $F_{\nu} \propto \nu^{\beta}$. 

To model the full spectral energy distribution from UVOIR to X-rays, we use the irradiated disc model \textsc{optxrplir} \citep{shidatsu16,kimura19}. The model assumes that the irradiating X-ray flux scales with the radius as $r^{-12/7}$ and that the irradiating flux consists of the canonical thermal disc contribution, an additional soft thermal Comptonization component, and a non-thermal (power law) component. The non-thermal component is assumed to be located close to the compact object from the innermost stable orbit to a radius $R_{\rm pl}$, while the soft thermal Comptonization component is located between $R_{\rm pl}$ and a radius $R_{\rm comp}$. The non-thermal component is parametrized with a power law index $\Gamma_{\rm pl}$, while the thermal Comptonization component is controlled by an electron temperature $kT_{e}$ and an optical depth $\tau$. The irradiated, multicolor disc is  characterised by an inner disc temperature $kT_{\rm bb}$, and is assumed to be located between $R_{\rm comp}$ and the outer disc edge, $R_{\rm out}$. 

Since the orbital parameters of the system are quite well determined, we restricted the outer disc radius to lie between the circularization radius
\begin{equation}
\frac{R_{\rm circ}}{a} = 0.074 \, \Bigg( \frac{1+q}{q^2} \Bigg)^{0.24} \approx 0.27, 
\end{equation}
\noindent and tidal radius 
\begin{equation}
\frac{R_{\rm tidal}}{a} = 0.112 + \frac{0.27}{1+q} +  \frac{0.239}{(1+q)^{2}} \approx 0.57.
\end{equation}
Here, $q = M_{2} / M_{\rm BH} = 0.072\pm$0.012 \citep[][Table \ref{system_param}]{torres20} is the mass ratio and $a \sim 4.5 \times 10^{11}$ cm is the orbital separation. The resulting limits on the outer disc radius are $10^{4.8} r_g \leq R_{\rm out} \leq
10^{5.4} r_g$, where $r_g$ is the gravitational radius. 

The irradiation efficiency in the \textsc{optxrplir} model is parameterized as the fraction of illuminating flux thermalized in the outer disc. This is given by $F_{\mathrm{out}} = f_{\mathrm{out}} (1-a_{\mathrm{out}})$, where $f_{\mathrm{out}}$ is the height of the disc at the outer edge divided by the outer radius and $a_{\mathrm{out}}$ is the albedo of the outer disc. We fix the albedo to a value of $a_{\mathrm{out}}=0.9$, which corresponds to a highly ionized and reflective disc surface and is typical for soft-state XRBs \citep{vanparadijs94,jimenezgarate02,gierlinski09}. 

The normalisation of this disc model depends on several physical parameters: the mass of the black hole ($M_{\rm BH}$), the total luminosity ($L$), the black hole spin, the disc inclination ($i$), and the distance to the source ($d$). In our modelling, we adopt $M_{\rm BH} = 8.5 M_{\odot}$, $i = 65^\circ$,  and $d=3$ kpc, as shown in Table \ref{system_param}. In addition, we assume a non-spinning black hole \citep{guan21,zhao21}. We also adopted solar abundances for all elements taken from \citet{wilms00}.

When we applied this model to the X-ray data -- also allowing for interstellar absorption by adding \textsc{tbnew}\footnote{An updated version of \textsc{tbabs} \citep{wilms00}} as a multiplicative component -- we found that unacceptable residual remained in the soft X-ray region. We, therefore, added a (relatively) low-temperature and low-normalisation black body component ($kT_{\rm bb} \sim$0.2 keV).\footnote{We also experimented with letting the abundances of the absorption component vary instead of adding the black body component, but this did not produce good enough results.} This component could potentially arise from the reprocessed emission lines in the illuminated disc atmosphere or wind base (see Section \ref{sec:reprocess}). Thus, we fit the X-ray spectrum with the \textsc{ISIS} parameterization \textsc{constant} $\times$ \textsc{tbnew} $\times$ (\textsc{optxrplir}+\textsc{bbody}). This model yielded acceptable fits to the X-ray region.

We then extended the data range considered to include the longer-wavelength UVOT and X-Shooter data. As a starting point, we fitted the above model to the whole broadband spectrum, but now also including an interstellar reddening component with $E_{\rm B-V}$=0.2 (see Section \ref{redden}). However, this model could not be fully reproduce the X-Shooter continuum, primarily because it underpredicts the infrared flux. Since the observation is taken in the soft state, it is unlikely that the required extra emission component arises from a jet or a companion star. However, as we show below in Section \ref{sec:sim_results}, the reprocessing in a wind or an atmosphere above the disc can produce infrared continuum emission comparable to that from an irradiated outer disc. 

In order to first obtain a rough approximation of the overall spectrum, we added a broken power-law component to account for the infrared flux deficit. We fixed the break frequency to 0.04 keV and the power law photon index after the break to $\Gamma=5$, leaving the photon index before the break and the normalisation as free parameters. Thus, the final model fitted to the overall broad-band SED is \textsc{constant} $\times$ \textsc{redden} $\times$ \textsc{tbnew} $\times$ (\textsc{optxrplir}+\textsc{bknpower}+ \textsc{bbody}). The best-fit model is shown in Fig. \ref{bbcont}, and the parameters are tabulated in Table \ref{params}. 

The above model produced a good fit ($\chi_{\mathrm{red}}^{2} \sim 0.9$) with the largest residuals arising in the very soft end of the \nicer\/ spectrum and in the optical and UV region. The soft X-ray residuals could arise either from a calibration issue, insufficient modeling of the soft X-ray absorption, or the effect of additional absorption/emission lines. The fitted value of the interstellar absorption is N$_{\mathrm{H}}$$\sim$1.3$\times$10$^{21}$ cm$^{-2}$, though, in line with the estimates from other methods described in Section \ref{redden}. The residuals in the optical/UV most likely arise from the insufficient modeling of the Paschen edge and possible differences to the assumed extinction law (\textsc{redden} uses the extinction law from \citet{cardelli89}).

\section{Modelling the Emission Line Spectrum}

In order to gain insight into the nature of the gas that produces the emission features in the X-Shooter spectrum of \source\/, we have modelled the spectrum with the Monte Carlo radiative transfer (MCRT) and ionisation code, \textsc{python}. 
The code and the kinematic model it uses are described in more detail in Section~\ref{sec:mcrt}, and the results of our modelling effort are described in Section~\ref{sec:sim_results}. 

As we shall see, the dense, cool, and quasi-static base of a thermally-driven disc wind can provide a suitable environment for forming the observed optical spectrum. However, it is important to note from the outset that we are {\em not} claiming that the presence of the wind is crucial to our ability to reproduce the observations. Instead, {\em any} irradiated layer above the disc with physical conditions similar to those we find in our wind base would likely produce a similar spectrum. 

Our choice to describe this region as the base of a thermally-driven wind is based on two considerations. First, such a wind is theoretically expected to be present in any system where a sufficiently large disc is subject to strong irradiation \citep[e.g.][]{begelman83,woods96,proga02,higginbottom19}, and -- as discussed in Section~\ref{introduction} -- observational signatures of disc winds have indeed been seen in many luminous XRBs. It is, therefore, natural to ask whether/how such a wind might also produce the observed optical emission lines. Second, \textsc{python} is a code designed to model outflows. Thus the density and velocity fields across the numerical grid are (usually) set up by specifying the parameters of a simple biconical disc wind (e.g., the mass-loss rate, the outflow opening angles, and the poloidal velocity law).

\subsection{Monte Carlo radiative transfer modeling}
\label{sec:mcrt}

\textsc{python} was originally developed by \cite{long02}\footnote{\textsc{python} and is a collaborative open-source project available at \url{https://github.com/agnwinds/python}.}. It has also been used by \cite{higginbottom18,higginbottom20} in their radiation-hydrodynamic simulations of thermally-driven disc winds in XRBs. Since the code was originally described, significant improvements have been made over the years by, among others, \cite{sim05} and \cite{matthews15}. We adopt the hybrid macro-atom scheme originally described by \cite{matthews15} for the models used here. Hydrogen and helium are treated in detail using the  macro-atom formalism developed by \cite{lucy02,lucy03}. Metals are treated as ``simple atoms'' in which a two-level atom approximation is used to treat line interactions, and the treatment of the bound-free continuum is also simplified.  Line transfer is treated in the Sobolev approximation, and strict radiative equilibrium (apart from a small cooling effect due to adiabatic cooling) is enforced in the co-moving frame; as a result, all the reprocessed line and continuum emission discussed in Section~\ref{sec:sim_results} has its energetic origins in the input radiation sources (in this case, the accretion disc).

Our MCRT simulations assume an axisymmetric wind structure with reflective symmetry about the disc plane and cylindrical symmetry about the $z$ axis; however, MC photon packets are free to propagate in 3D. The density structure and velocity field throughout this axisymmetric outflow are determined using the kinematic prescription described below in Section~\ref{sec:wind_model}. Once the density and velocity grid is specified, we aim to determine the effect of the outflow on the observed spectrum by using \textsc{python} to carry out the ionization and radiative transfer calculations. 
Detailed descriptions of how this works can be found in the references provided above. However, briefly, the calculations are carried out in two distinct stages. First, \textsc{python} generates sets of photon packets representing the radiation of all relevant system components over the full range of frequencies. Since we are only concerned with the soft state, the irradiated disc is currently the only component we consider in these simulations. These photon packets are followed as they make their way through the outflow and interact with the wind material. During this process, the ionisation and temperature structure of the wind is iteratively updated until it converges. At this point, the ionisation and temperature structures are frozen, and the calculation proceeds to the second stage, in which a detailed spectrum is calculated. Only photon packets in the spectral range of interest are generated at this stage.

\subsubsection{Radiation Sources}
\label{sec:radiation_sources}

\begin{figure}
 \centering
 \includegraphics[width=\linewidth]{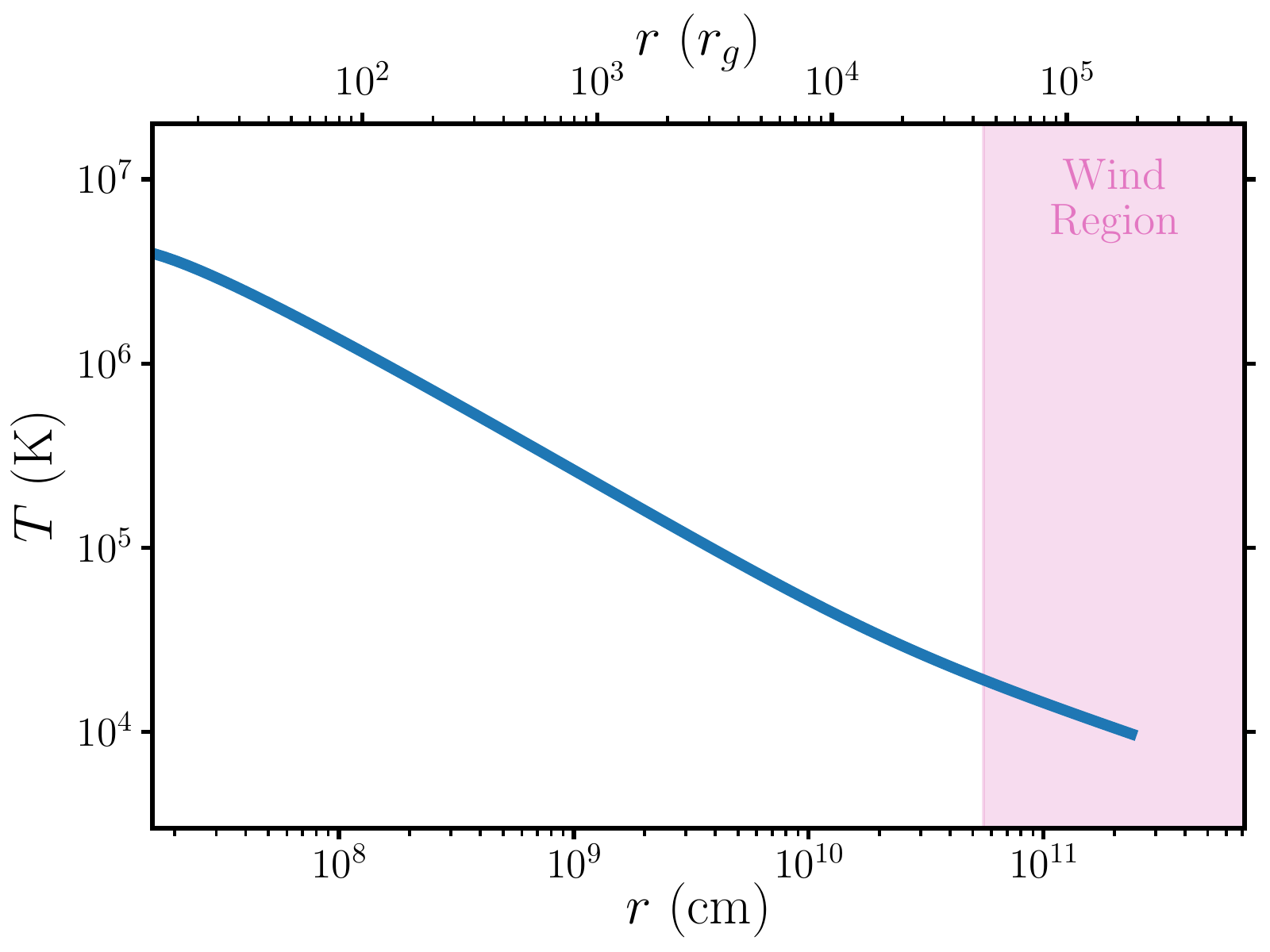}
 \caption{The temperature of the disc as a function of radius in our model. The pink-shaded region marks the approximate location of the disc wind, which has streamlines extending beyond the disc's outer edge.}
 \label{fig_disc_temp}
\end{figure}

In order to model the X-shooter spectra of \source, we assume that the accretion disc is optically thick and geometrically thin. Its radiation field is then approximated by an ensemble of concentric annuli that emit blackbody radiation at the local disc temperature. The temperature profile we use is that suggested by the best-fitting full SED model described in Section~3.3. In the inner regions, viscous dissipation dominates, and $T_{\rm visc}(r) \propto r^{-3/4}$; in the outer regions, irradiation dominates, flattening the temperature distribution to $T_{\rm irr}(r) \propto r^{-3/7}$. The overall profile can be described by the sum $T^4(r) = T_{\rm visc}^4(r) + T_{\rm irr}^4(r)$, where 
\begin{equation}
T^4_{\rm visc}(r) = \frac{3 G M_{\rm BH} \dot{M}}{8 \pi \sigma_{\rm SB} r^3},
\end{equation}
and 
\begin{equation}
T_{\rm irr}^4(r) = \mathcal{C}(r) \frac{\dot{M} c^2}{4\pi \sigma_{\rm SB} r^2}.
\end{equation}
Here, $G$ is the gravitational constant, $\dot{M}$ is the mass accretion rate, $\sigma_{SB}$ is the Stefan-Boltzmann constant, and $\mathcal{C}(r)$ is the fraction of the luminosity ($L_{\rm X}=\dot{M} c^2$) thermalised in the disc depending on the disc scale height and albedo. \textsc{optxrplir} parameterises this as $\mathcal{C}(r)=(2/7) f_{\rm out} (1-a_{\rm out}) (r/r_{\rm out})^{2/7}$, with the parameters and their values described in Section~3.3. The resulting temperature distribution is shown in Fig. \ref{fig_disc_temp}.

\subsubsection{Biconical Wind Model}
\label{sec:wind_model}
We parameterise the wind as a biconical flow according to a prescription developed by \cite{shlosman93}. We use this kinematic model because it is flexible and fairly simple and because we have developed a certain amount of insight into adjusting its parameters to produce synthetic spectra that resemble observations. 

The \cite{shlosman93} model defines the wind in terms of streamlines that arise from the disc between the limiting radii $r_{\rm min}$ and $r_{\rm max}$, with angles relative to the disc normal given by 
\begin{equation}
\theta(r) = \theta_{\rm min}+ (\theta_{\rm max}-\theta_{\rm min}) \left (  \frac{r-r_{\rm min}}{r_{\rm max}-r_{\rm min}} \right ) ^{\gamma}.
\end{equation}
Here, $\theta_{\rm min}$ and $\theta_{\rm max}$ are the angles of the streamlines with respect to the polar axis at the inner and outer edge of the wind, and the exponent $\gamma$ controls the concentration of streamlines. In our modelling here, we limit ourselves to the case $\gamma = 1$, so the streamlines are not concentrated toward the inner or outer edge of the flow. 

Material flows along the streamlines according to a prescription that splits the velocity field into poloidal and azimuthal components.  The poloidal velocity at a poloidal distance $l$ along a streamline is given by
\begin{equation}
v_l (l) = v_0 + (v_{\infty}-v_0) \left ( \frac{(l/R_v)^{\alpha}}{1+ (l/R_v)^{\alpha} } \right ),
\end{equation}
where $R_v$ is an acceleration length (the point at which one reaches half the final velocity), $v_0$ is the initial velocity, $v_{\infty}$, is the terminal velocity, and $\alpha$ determines the sharpness of the transition to high velocity. The azimuthal velocity, $v_\phi$ is set so that material in the wind initially rotates with the (Keplerian) disc but then conserves specific angular momentum. Thus, far from the disc's rotation axis, the wind's rotational velocity component, $v_\phi$, is small.

Finally, to complete the description, one must define the mass outflow rate as a function of position.  In the \cite{shlosman93} parameterization, the mass-loss rate per unit surface area, $\dot{m}$, is assumed to be given by 
\begin{equation}
\dot{m} \propto \dot{M}_{\rm wind}  r^{\lambda} \cos \left(\theta(r) \right).
\end{equation}
Here, $\dot{M}_{\rm wind}$ is the total mass-loss rate carried by the outflow.  For a narrow flow with $\theta_{\rm min}$ not very different from $\theta_{\rm max}$, an exponent of $\lambda \simeq -2$ results in constant mass loss rate as a function of the disc radius $r$.  For the model, we ultimately found that for \source, where $\theta_{\rm max}$, is close to 90\degr, mass loss is concentrated toward $r_{\rm min}$, because of the cosine term in the equation above.

\subsubsection{Arriving at a fiducial model}

\label{sec:arriving_fiducial}

Even though the \cite{shlosman93} prescription for an accretion disc wind is fairly simple, it involves quite a few tunable parameters. Given that the MCRT calculations require substantial computational resources, it is impractical to create a grid of models and to search a vast parameter space to find a `best-fit' model to an observed spectrum. Instead, our approach was to iterate from a set of parameters we consider physically plausible, adjusting one or two parameters at a time until we obtained spectra that look qualitatively similar to the observed spectrum of \source\ at about the proper distance and inclination. 

The main characteristics of this model are inspired by recent radiation-hydrodynamical simulations of thermally driven disc winds in XRBs \citep[e.g.][]{higginbottom15, shidatsu16, higginbottom18, shidatsu19, higginbottom20}. Thus, the outflow only begins at $r_{\rm min} \simeq 10^5~R_{g}$, since thermal driving is only effective relatively far out in the disc (where the inverse Compton temperature -- which is reached in the uppermost layers of the irradiated disc atmosphere -- can exceed the escape speed). The adopted mass-loss rate in the model is $\dot{M}_{\rm wind} = 10^{-8}~M_\odot~{\rm yr}^{-1}$, which corresponds to about half of the accretion rate. This, too, is not unreasonable, both empirically \citep[e.g.][]{ponti12} and for a thermally-driven wind \citep[e.g.][]{higginbottom15, shidatsu16, higginbottom18, shidatsu19, higginbottom20}. To match the equatorially-concentrated density structure of thermal disc winds, the inner and outer opening angles of the outflow have been set to  $60^\circ$ and $89.9^\circ$, respectively. However, the simulated spectra are fairly insensitive to this choice. The overall physical conditions throughout the fiducial model are discussed further in Section~\ref{sec:physical_conditions}.

To obtain spectra that resembled \source, we had to modify some of the wind parameters, particularly the initial velocity at the base of the streamlines, $v_0$, and the volume filling factor, $f_V$. Both of these parameters are important in dictating the density in the line-forming region and the overall reprocessing efficiency of the wind; however, they are also not unique in the sense that other parameters we did not adjust could have equal importance. We found we needed the material to be moderately clumpy, with a volume filling factor of $f_V=0.02$. The filling factor and mass-loss rate affect the results primarily via their ratio $\dot{M}_{\rm wind} / f_V$, which helps to set the density in the line-forming region (see Section~\ref{sec:line_forming}). Once we had arrived at a ``fiducial'' model, we then explored changes in some of the other parameters to verify that the results were not locally very sensitive to them. However, we have not attempted to go beyond a general agreement between the model and the observed data. We comment further on the plausibility of clumping in the wind and the overall feasibility of line formation in the transition region of a disc wind, in Section~\ref{sec:discuss_density}. 

\begin{table}
\centering
\caption{Parameters of the fiducial wind model.}\label{model_params}
\begin{tabular}{lrl}
\hline\hline
Parameter & Value & Comment \\
\hline
$\dot{M}_{\rm wind} $       &           $10^{-8}~M_{\odot}~{\rm yr}^{-1}$ &  Mass loss rate \\
$r_{\rm min}$                    &         $8 900~r_g$ &  Inner edge of wind at base \\
$r_{\rm max}$                  &           $12 700~r_g$ & Outer edge of wind at base  \\
$\theta_{\rm min}$           &         $60\degr$ & Angle of flow at inner wind edge \\
$\theta_{\rm max}    $     &        $89.9\degr$ & Angle of flow at outer wind edge \\
$\lambda$                &             $-1$ & Mass loss radial exponent\\
$v_0$                      &                         $1000~{\rm cm~s}^{-1}$ &  Velocity at base \\
$v_{\infty}(r)$             &    $1~v_{\rm esc}(r)$ & Terminal velocity \\
$R_v $             &            $5\times10^{11}$ cm & Acceleration length\\
$\alpha$               &           $2$ & Acceleration exponent\\
$\gamma$           &     $1$ & Streamline skew \\
$f_V$ &   $0.02$ & Filling factor\\
\hline 
\end{tabular}
\end{table}

\begin{figure*}
 \centering
 \includegraphics[width=0.95\linewidth]{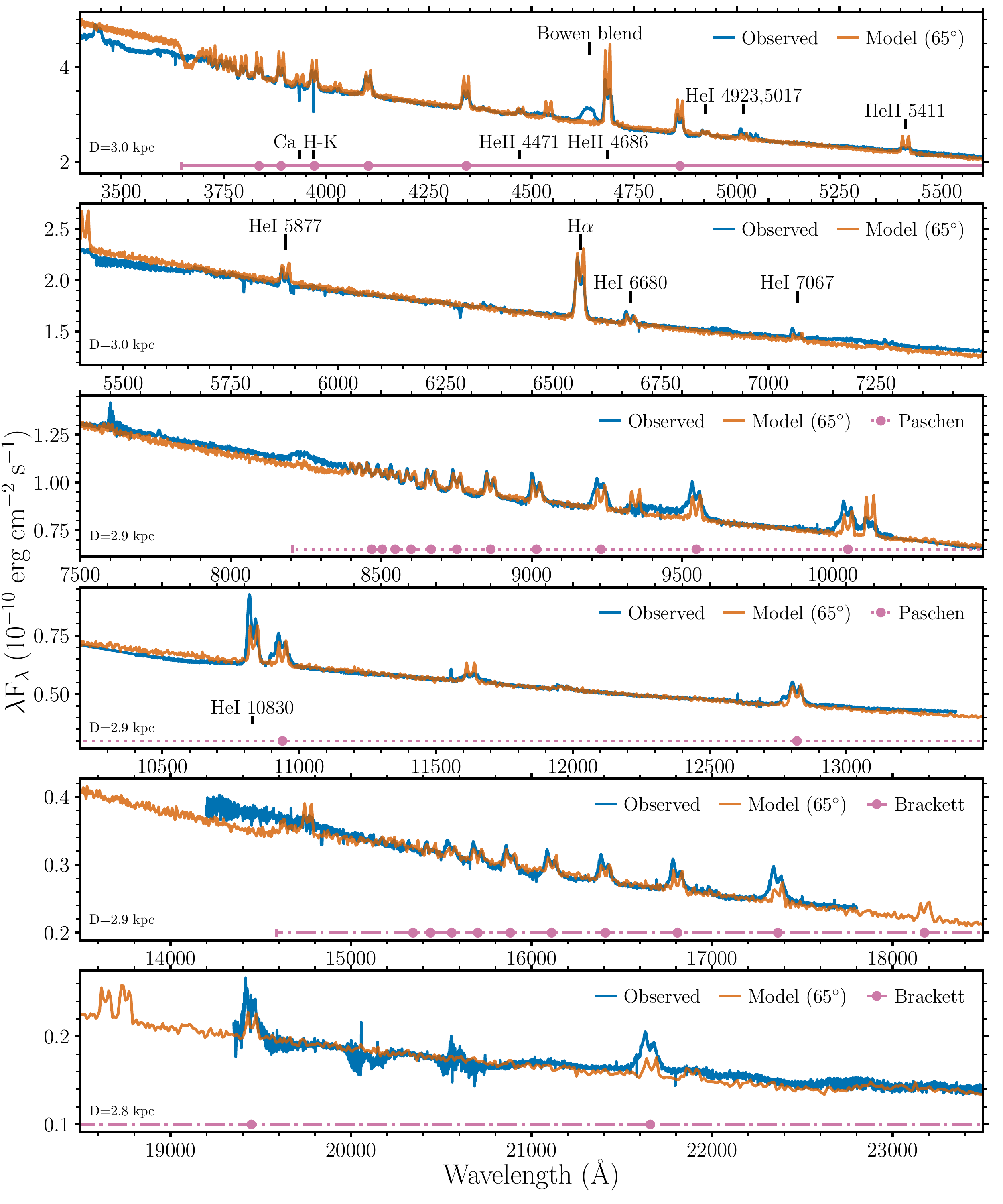}
 \caption{A comparison of the synthetic spectrum generated with our MCRT modelling to the observed spectrum of \source. The synthetic spectra have been renormalised slightly to match the continuum and require a slightly different normalisation at different wavelengths, reflecting a slight disagreement with the observed continuum shape. The prominent emission lines are marked, as are the Brackett and Paschen series. Except for the Bowen blend and higher-order lines not captured by our atomic models, the wind model successfully produces {\em all} of the emission lines seen in the spectrum with similar shapes, equivalent widths, line widths, and relative intensities. There are some notable differences in the asymmetries of the emission lines, which are discussed in the text.
 }
 \label{fig:python_spectrum}
\end{figure*}

\subsection{Simulation Results}

\label{sec:sim_results}

\subsubsection{Emission Line Spectrum} \label{sec:sim_results_line}
In Fig.\ \ref{fig:python_spectrum}, we compare our fiducial model to the observed spectrum of \source. Since we are primarily interested in the emission line and photo-ionisation edge signatures, we renormalise the model spectrum to match the data in each panel of the figure, i.e., in each wavelength region. The distances implied by this renormalisation are indicated in the figure. 

At a qualitative level, the simulated spectrum resembles that of \source\  at the time of the observations using the disc temperature profile that was generated to fit the multiwavelength SED. Most of the rich sets of  double-peaked lines in the observed source also appear in the simulated spectrum. The normalisation of the simulated spectrum correspond to a source at about the correct distance and inclination and the optical/IR continuum has about the correct slope due to a substantial contribution by free-free and bound-free emission from the wind region. The model spectrum approximately matches the strengths and widths of the observed hydrogen and helium features and the H and K lines of calcium. The most prominent lines missing from our model are the O~\textsc{iii} and N~\textsc{iii} Bowen lines near $\lambda$3450 and $\lambda$4650. These features are produced by fluorescence mechanisms not currently included in {\sc Python}. 

Although the lines in both the observed and model spectra are double-peaked and of comparable width, there are interesting differences in the detailed line shapes; on the whole, the model lines seem to be fairly symmetric or to have stronger red than blue peaks, whereas the observed spectrum has lines in which the blue peak is brighter than the red peak. The asymmetry in the model line profile is straightforward to understand. Most of the observed spectrum is generated by photons from a relatively thin layer above the near side of the disc; radiation associated with the back side of the disc is blocked by the disc itself. Photons produced in this layer can interact with material downstream in the outflow, moving toward the observer. As a result, any bound-bound absorption or scattering occurring in the outflow will preferentially affect the blue wings of the lines. If the spectrum produced by the emitting layer were flat or the line opacity higher, this would produce the classic blueshifted absorption signature associated with wind-formed P-Cygni profiles. However, the spectrum incident on the wind already contains strong and broad emission lines, so the absorption and scattering results only in the suppression of the blue line wings.

There are several possible explanations why the observed spectra do not show the same character, the most obvious being that the outflow was not axisymmetric at the time of these observations. The Keplerian period at the outer edge of the disc is almost 6 hours, considerably longer than the time over which the spectra reported here were obtained. Given the relatively low velocities implied by the model we are using to simulate the spectra, it is also not clear that all of the wind is outflowing at all times. If we observe the system through gas settling onto the disc, one might expect the red peak to be stronger than the blue peak. In any event, we note that other spectra have been taken of \source\ in the same spectral state, showing brighter red peaks \citep{sanchezsierras20,munozdarias19}, so it is clear that the asymmetries are time variable.

\subsubsection{Continuum Emission and Wind Reprocessing} \label{sec:reprocess}
The overall slope of the spectrum shown in Fig.~\ref{fig:python_spectrum} roughly matches the observed continuum due to a significant amount of emission that arises in the wind, especially at the longer wavelengths. The model somewhat underpredicts the NIR flux, as indicated by the lower implied distance in this band. However, in the absence of reprocessed wind emission, the disc spectrum would have an even steeper spectral slope and underpredict the NIR flux more severely. 

\begin{figure}
 \centering
 \includegraphics[width=0.95\linewidth]{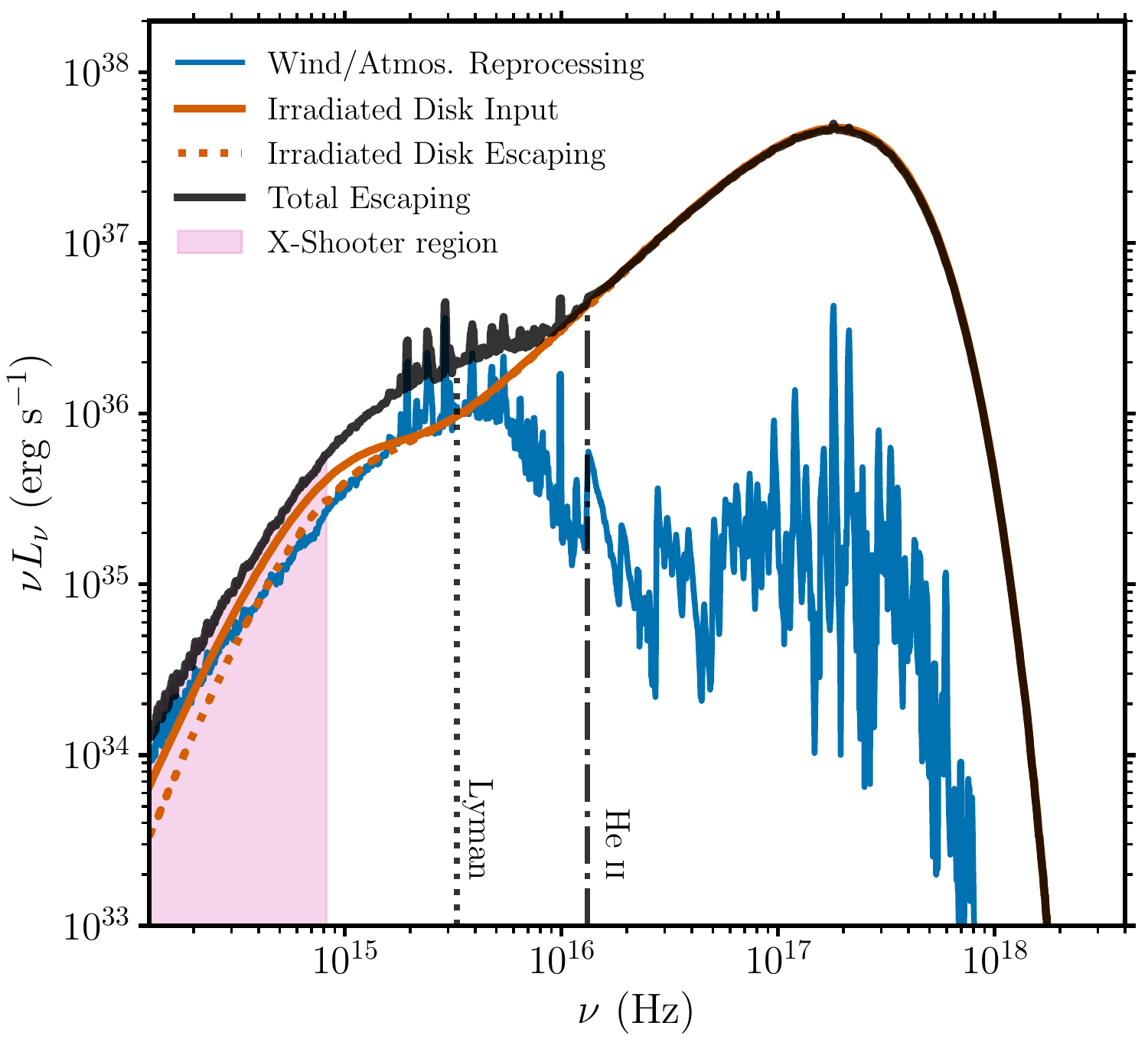}
 \caption{The broadband spectral energy distribution obtained from the wind modelling and binned over all viewing angles. The total emergent spectrum is shown in black, the input irradiated disc spectrum in orange, and the spectrum produced by reprocessing in the wind in blue. The He~\textsc{ii} and Lyman ionisation edges are marked, and so is the frequency region covered by the X-Shooter observations. The figure illustrates how the wind reprocesses the input radiation into continuum and line emission. Prominent recombination edges can be seen in the wind spectrum. The overall energetic output of the wind is small compared to the bolometric, but due to the spectral shape, this reprocessing still has a profound impact on the optical spectrum.}
 \label{fig:python_sed}
\end{figure}

Wind reprocessing is an important factor in determining the emergent continuum in our modelling; this effect can be seen in Fig.~\ref{fig:python_sed}, which shows the total broadband SED escaping from our MCRT simulation, averaged over all viewing angles. We also show the input disc SED and escaping contributions from the disc and wind reprocessing. The escaping spectrum in the extreme-UV (EUV) and soft X-ray is largely unmodified, and the energy budget is dominated by the thermal bump from the hot irradiated disc, which peaks at $\approx 3 \times 10^{17}~{\rm Hz}$ as in Fig.~\ref{bbcont}. However, while the wind has a modest effect on the overall energetic output, the impact on the spectrum redward of the He~\textsc{ii} and Lyman edges -- including in the X-Shooter region -- is much more dramatic. The wind reprocesses the hot disc radiation into a quasi-thermal optical to near-infrared bump. Interestingly, the wind-processed continuum dominates over the disc radiation at the red end of the X-Shooter spectral region. In total, the wind absorbs around $3\times 10^{36}~{\rm erg~s}^{-1}$ of the total luminosity, most of which is absorbed close to the peak of the SED, leading to a slight suppression of the X-rays and EUV which can barely be seen on the figure. This absorbed energy is radiated through a series of strong emission lines and recombination continua, and the wind noticeably changes the continuum slope in multiple wavebands. Thus, Fig.~\ref{fig:python_sed} shows that dense disc winds or disc atmospheres can have a dramatic impact on optical, UV, and NIR continua in XRBs, and possibly also in other accreting systems \citep[see, e.g.,][]{matthews15,parkinson22}.

Inspection of Fig.~\ref{fig:python_spectrum} shows that the shape of the continuum is not quite right close to the Balmer, Paschen, and Bracket limits (marked with vertical lines).  Some of this effect, particularly at the Balmer limit, may be due to photo-absorption within the wind but could also be due to the limited number of levels used in our atomic model for hydrogen (up to a principal quantum number of 20). As a result, we do not include finely-spaced lines very close to the ionisation edge, which can form a pseudo-continuum leading up to the relevant limit. In particularly, the model includes 18 lines in the Balmer series, 17 in the Paschen series, and 16 in the Brackett series, which implies that regions of widths 36 \AA, 185 \AA, and 241\AA \ redward of the Balmer, Paschen, and Brackett limits, respectively, are not modelled accurately. In addition, in some of the included high-order lines the line strengths will not be entirely accurate because radiative cascades from levels just above them are not included.

\subsubsection{Physical Properties of the Fiducial Wind Model}
\label{sec:physical_conditions}

\begin{figure*}
 \centering
\includegraphics[width=0.98\textwidth]{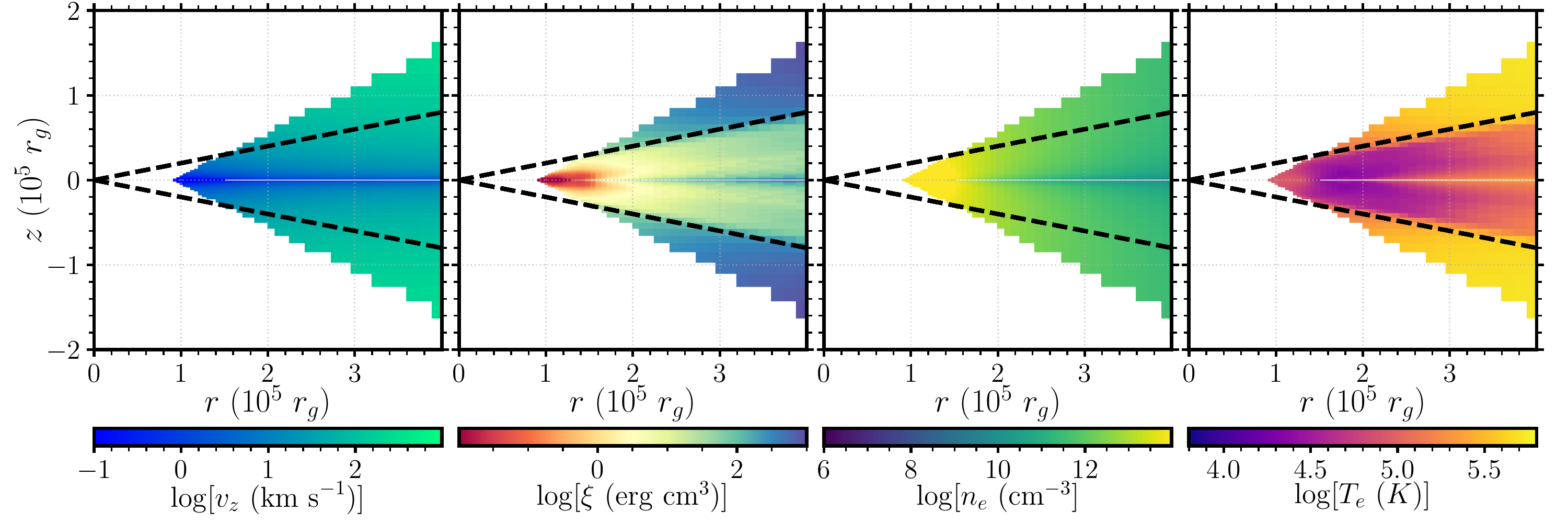}
 \caption{The physical conditions in the adopted wind model. The panels show, left to right, colour maps of logarithms of the vertical velocity component, $v_z$, the ionisation parameter $\xi$, the electron density, $n_e$, and the electron temperature, $T_e$, for the wind used to produce the synthetic spectra shown in Fig.\ \ref{fig:python_spectrum}. The black dashed line shows a wedge with a scale height of $z/r=0.2$. Only the region inside the disc radius is shown, but this is the source of essentially all of the line emission, and only the two positive $r\geq0$ quadrants are plotted.}
 \label{fig:wind}
\end{figure*}

The overall physical properties throughout the fiducial model are illustrated in Fig.\ \ref{fig:wind}, which shows the vertical wind velocity, $v_z$, ionisation parameter, $\xi$, electron density, $n_e$ and electron temperature, $T_e$, in the converged simulation. Fig.\ \ref{fig:wind} demonstrates some of the overall characteristics of this roughly equatorial wind model. The wind base is dense, and a wedge-shaped region is formed, with aspect ratio $z/r\approx 0.2$, of moderate ionization material at temperatures conducive to forming hydrogen and helium recombination lines ($T_e \simeq 2-4~\times10^4{\rm K}$). The wind base is also subsonic, giving way to higher velocity, lower density material further out. In addition, the upper layers of the wind are much hotter than the wind base, reaching maximum temperatures of $T_e \simeq 5\times10^5~{\rm K}$.

In this type of wind model, the only natural location for forming optical lines is the base of the outflow, i.e., the interface between the relatively cool disc material close to the mid-plane and the X-ray-heated material at the top of the atmosphere. The transition region between these regimes is typically characterized by thermal instability, i.e., material moving upwards in the atmosphere tends to heat up extremely rapidly as it does so. Therefore, conditions suitable for forming optical lines are likely to be found in a fairly thin layer and relatively close to the mid-plane. As we shall see below (see Section~\ref{sec:line_forming}), this is where most of the optical line formation takes place in our model. 

However, it is important to remember that identifying the optical line-forming with, specifically, the base of a thermally-driven wind is certainly not unique. {\em Any} layer above the disc plane with conditions similar to those found in our modelling would likely produce a similar optical emission line spectrum. This issue is discussed further in Section~\ref{sec:discuss_density}.

\subsubsection{The Line-Forming Region}
\label{sec:line_forming}
\begin{figure*}
 \centering
\includegraphics[width=0.98\textwidth]{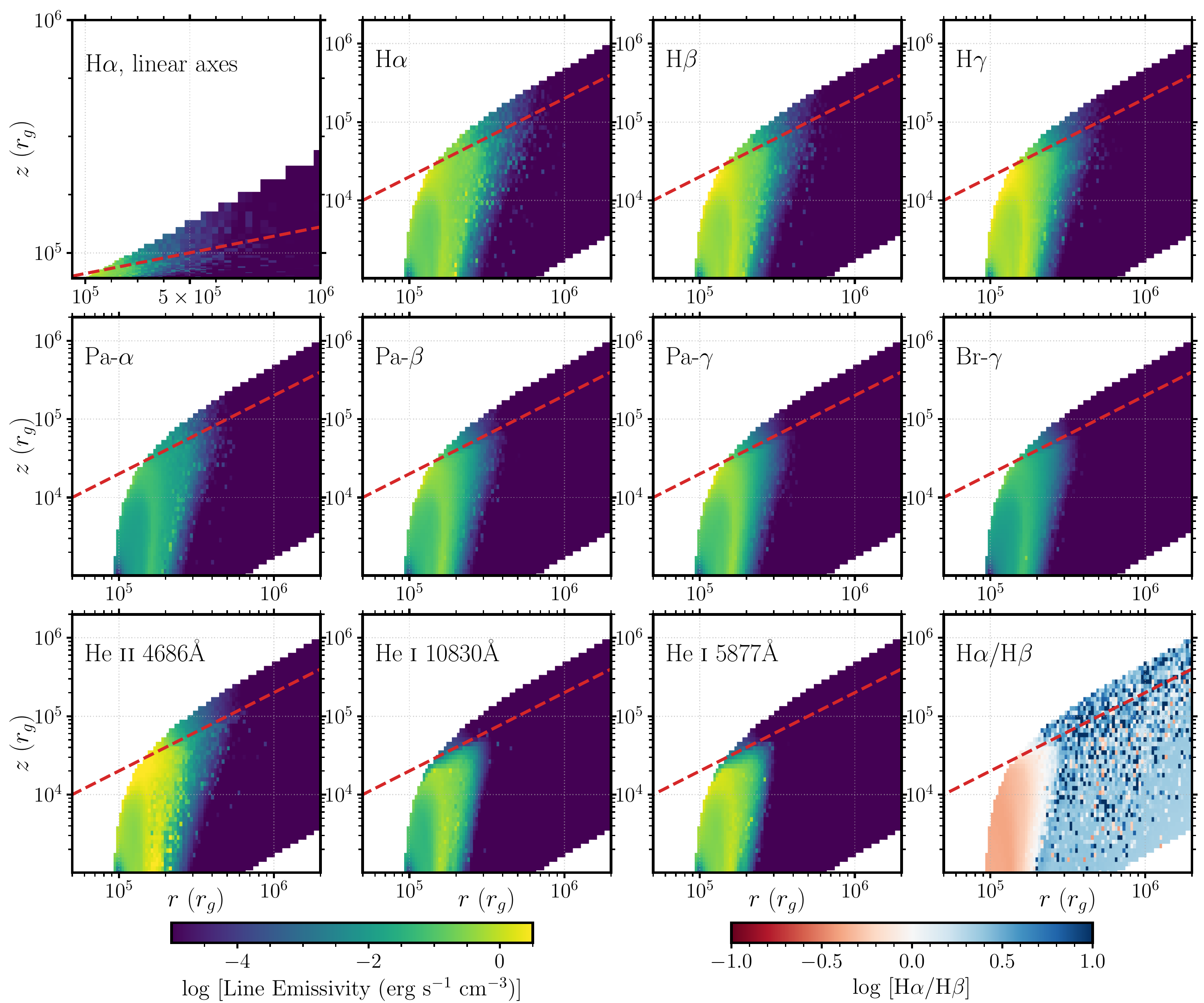}
 \caption{Various line emissivities throughout the wind. Line emissivities in each wind grid cell are plotted for some of the more prominent hydrogen and helium lines on a logarithmic spatial scale, except for the top left panel, which shows a linear scale. The linear scale is plotted for easy comparison with Fig.~\ref{fig:wind} and to show the disc-like, flattened geometry of the line formation region. The red dashed line marks $z/r=0.2$. The bottom-right panel shows the ${\rm H}\alpha/{\rm H}\beta$ line ratio in the wind, which is relatively low ($\lesssim 1$) in the base of the wind but increases outwards as the density drops.}
 \label{fig:lines}
\end{figure*}

In  Fig.\ \ref{fig:lines}, we show the spatial distribution of line emissivities (as well as the H$\alpha$/H$\beta$ ratio discussed in Section~\ref{sec:balmer_dec}) of some of the more prominent lines present in the modelled spectrum. The line emissivity here is the {\em local} line emissivity in a grid cell, in the sense that it includes, as a multiplicative factor, the Sobolev escape probability, but this does not mean this line emission necessarily escapes to the observer. We show these emissivities on logarithmic axes to show sufficient detail, but also show a reference plot on linear axes for comparison with Fig.\ \ref{fig:wind}. This choice of presentation is necessary because the line emission is produced relatively close to the disc. For example, $90\%$ of the H$\beta$ emission is confined to $z<10^{4}~r_g~(z<5\times 10^{10}~{\rm cm})$ and $z/r\lesssim 0.2$. Given the position of the emission, it is not surprising that the widths of the lines are very similar to the projected rotational velocity of the disc, which, for reference, is 625 $\rm km\:s^{-1}$ at the point where the X-ray illuminated disc ends.

The line formation region is inevitably stratified in terms of its physical properties (as is apparent from Figs.\ \ref{fig:wind} and \ref{fig:lines}), but it is nonetheless instructive to consider some characteristic physical conditions. We, therefore, order the simulation cells by a given physical quantity ($n_e$, $T_e$, and $v_z$) and calculate the cumulative distribution function (CDF) of the H$\beta$ line luminosity in each case; the median (${\rm CDF} = 0.5$) value then gives us a reasonable characteristic value. This exercise reveals characteristic conditions in the H$\beta$ line-forming region of $n_e \sim 10^{13}~{\rm cm}^{-3}$, $T_e \sim 40,000~{\rm K}$ and $v_z\sim10~{\rm km~s}^{-1}$. This region can therefore be broadly characterised as a fairly thin wedge of dense, warm gas moving at subsonic to transonic velocities and is perhaps more accurately described as an atmosphere than a wind (see discussion in Section \ref{sec:discuss_density}).

\subsubsection{The Balmer decrement}
\label{sec:balmer_dec}

The Balmer decrement describes the decreasing ratios of line intensities in the Balmer series, typically defined with respect to the intensity of H$\beta$ \citep{Burbidge1953,osterbrock}. The Balmer decrement can be used to probe the physics of the line-emitting region by comparing the observed value to theoretical expectations. We calculated the Balmer decrement in \source\ from the continuum-subtracted line fluxes of H$\alpha$, H$\beta$, and H$\gamma$. The dereddened Balmer decrement in \source\ is rather flat, with ${\rm H}\alpha/{\rm H}\beta\approx$1.3 and ${\rm H}\gamma/{\rm H}\beta\approx0.95$. These values are significantly lower than the canonical value of ${\rm H}\alpha/{\rm H}\beta \approx 3$ expected from case B recombination theory \citep[e.g.][]{osterbrock}, and are more in line with case A recombination. These values also correspond to an emitting region temperature 10000 K and a density of 2.5$\times$10$^{12}$ cm$^{-3}$ using the model of \citet{adams74}.  

Flat Balmer decrements have been observed in various astrophysical sources with several explanations proposed for their origin. For example, \cite{williams1980} find a flat Balmer decrement in CVs, proposing emission from a dense region near the disc, and similar proposals have been made for tidal disruption events \citep{short2020}. In addition, a series of authors have calculated theoretical curves with varying assumptions about atomic physics and considering a range of physical conditions (density, temperature, optical depth, and ionizing radiation field). While the detailed behaviour depends on these parameters, the general trend is that the ${\rm H}\alpha/{\rm H}\beta$ can initially increase somewhat with density before decreasing to $\lesssim 1$ above $n_e \sim 10^{12}-10^{13}~{\rm cm}^{-3}$ \citep{adams74,drake1980,williams1980}. At these high densities, the lines can become optically thick, collisional de-excitations become important, and the ${\rm H}\alpha/{\rm H}\beta$ ratio is no longer set by the recombination rates. Instead, the level populations tend towards the local thermodynamic equilibrium (LTE) values at the local $T_e$. One caveat here is that radiative processes can also affect the Balmer decrement. For example, \cite{elitzur1984} highlights the importance of induced processes in intense radiation fields. In contrast, \cite{ferland1982} shows that flat Balmer decrements can be produced partly by stimulated emission, which drives the populations to LTE at lower densities. 

\begin{figure}
 \centering
\includegraphics[width=\linewidth]{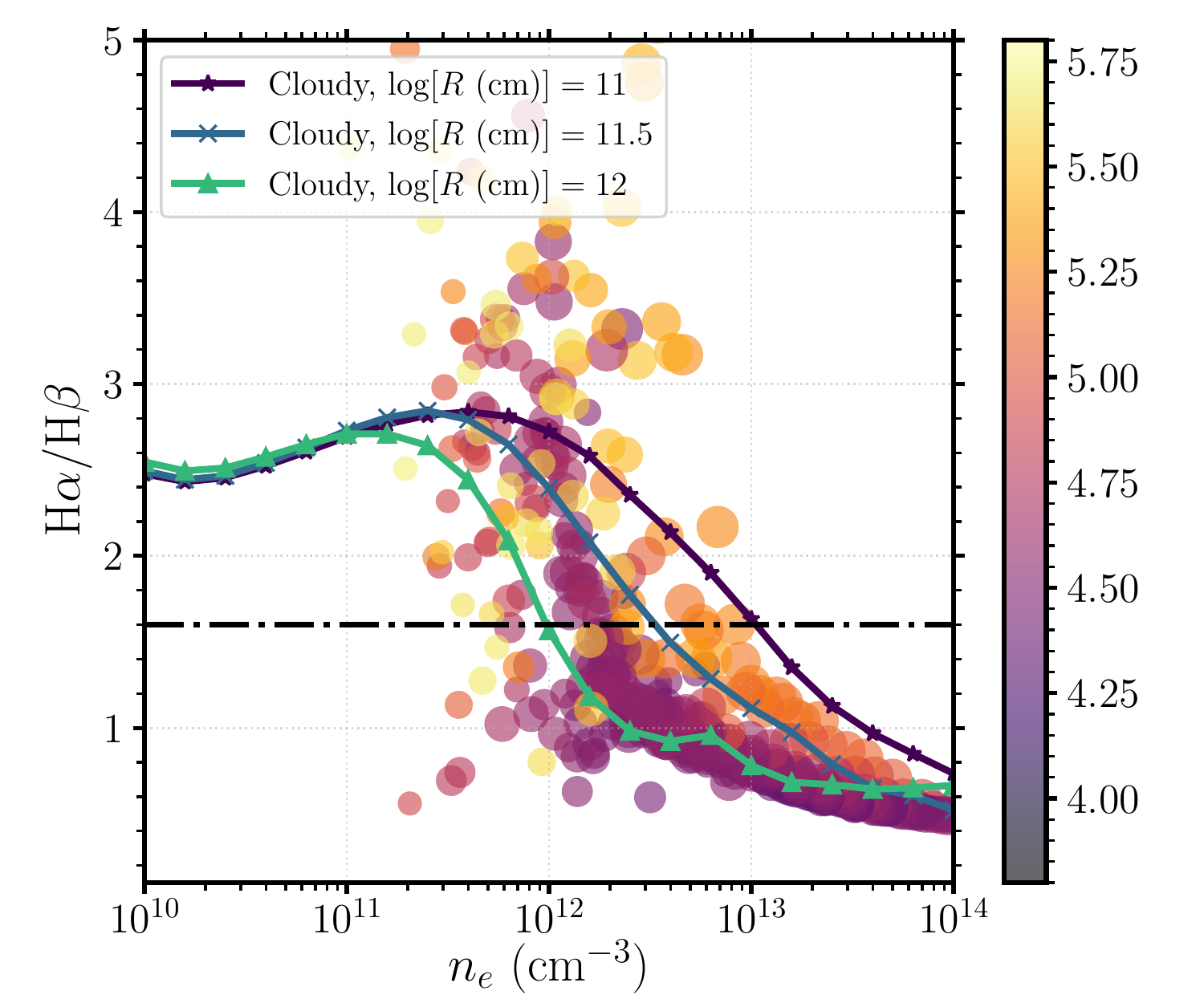}
 \caption{The Balmer decrement (${\rm H}\alpha/{\rm H}\beta$ ratio) from our photoionisation and radiative transfer modelling. The ${\rm H}\alpha/{\rm H}\beta$ ratio is plotted as a circle for every wind grid cell in the MCRT simulation, colour-coded by $\log[T_e~(K)]$. Three curves of ${\rm H}\alpha/{\rm H}\beta$ are overlaid from the Cloudy simulations described in the text, and the dot-dashed line marks the inferred value of ${\rm H}\alpha/{\rm H}\beta=1.3$ from the observations for $E_{\rm B-V}=0.2$. Both codes independently find that $n_e \gtrsim 10^{12}~{\rm cm}^{-3}$ is needed to produce the observed flat Balmer decrement. However, material over a range of densities contributes to the line emission in the MCRT simulation.}
\label{fig:balmer_decrement}
\end{figure}

To gain more insight and to examine the behaviour of the ${\rm H}\alpha/{\rm H}\beta$ ratio  when the stratified wind structure is illuminated by a SED appropriate for \source, we can examine the Balmer decrement from our MCRT modelling. In the bottom right-hand panel of Fig.~\ref{fig:lines}, we show the ${\rm H}\alpha/{\rm H}\beta$ ratio throughout the wind. ${\rm H}\alpha/{\rm H}\beta \lesssim 1$ in the dense base of the wind and throughout the majority of the line formation region. Higher up in the flow, the ratio increases, but we can also see the effect of Monte Carlo noise.  In Fig.~\ref{fig:balmer_decrement}, we show the ${\rm H}\alpha/{\rm H}\beta$ ratio as a function of $n_e$ from our MCRT wind modelling. Each point represents an individual grid cell and is colour coded by temperature. Only cells that contribute at least $0.01$\% of the total H$\alpha$ luminosity are plotted to illustrate the characteristic values in the line formation region. Additionally, we overlay three curves calculated with the photoionisation code Cloudy \citep{ferland1998,ferland2017}. In the Cloudy simulations, we considered a cloud of gas with a fixed column density of $N_H = 10^{23}~{\rm cm}^{-2}$ illuminated by the same irradiated disc SED we use in our wind modelling. We placed the cloud at three different radii ($\log [r~({\rm cm})] \in {11,11.5,12}$) and calculated the ${\rm H}\alpha/{\rm H}\beta$ ratio at different densities $n_e$. The curves have a similar characteristic shape to the calculations presented in the literature mentioned above. However, the value of $n_e$ at which the low ${\rm H}\alpha/{\rm H}\beta$ ratio regime is reached depends somewhat on the radius of the cloud (and therefore the dilution of the radiation field). 

Our overall interpretation of the flat Balmer decrement in the emission line spectrum is similar to that of our MCRT wind modelling results. The flat Balmer decrement is consistent with emission from a dense wind base or disc atmosphere, a scenario that has also been suggested to explain the flat Balmer decrement in CVs \citep{williams1980} and TDEs \citep{short2020}. We discuss this scenario further in Section~\ref{sec:discuss_density}.

\section{Discussion} \label{discussion}

\subsection{Optical line formation in the cool base of an X-ray heated wind or corona}
\label{sec:discuss_density}

A thermally-driven outflow from an X-ray-heated accretion disc is not the most immediately promising environment for generating strong optical line emission. The conditions required to produce the observed hydrogen and helium recombination lines involve relatively low temperatures ($T \simeq 10^4$~K), moderate ionisation, and relatively high densities. By contrast, thermal driving only takes off once material in the atmosphere is rapidly heated to near the inverse Compton temperature ($T_{\rm IC} \simeq 10^6$~K~-~$10^7$~K), which happens at very high ionisation parameters. Thus the rapidly outflowing material is unlikely to be a source of significant optical emission (unless it is exceptionally clumpy and/or multiple phases co-exist in the outflow).

In the context of a thermally-driven wind model, this leaves the base of the outflow -- i.e., the interface between the cool, hydrostatic disc and the hot, outflowing wind -- as the most likely optical line-forming region. Are the properties of the line-forming region in our wind model consistent with such a scenario?

As already noted in Section~\ref{sec:arriving_fiducial}, both the mass-loss rate ($\dot{M}_{w} \simeq \dot{M}_{\rm acc}$) and the location of the outflow on the disc ($R_{\rm min} \simeq 10^5 R_{g}$) in our model are reasonable for a thermally-driven disc wind. We can estimate the density in the interface region between the hydrostatic disc and the outflow from the continuity equation. For simplicity, let us consider a quasi-spherical outflow starting at $R_0 \simeq R_{\rm min} \simeq 10^{11}$~cm with velocity $v_0 \simeq c_s$, where $c_s$ is the sound speed at the base of the interface region. Temperatures here are of order $10^4$~K, so we can write $c_s \simeq \sqrt{T/(10^4~{\rm K})}~{\rm km~s}^{-1}$. Mass continuity then implies an initial density of

\begin{equation}
    n_{0} \sim \frac{\dot{M}_{w}}{f_V f_{c} 4\pi R_{0}^2 m_{p} v_{0}},
\end{equation}
\noindent where $f_V$ is the filling factor, $f_{c}$ is the covering factor, and $\dot{M}_{w}$ is the mass loss rate of the wind. Plugging in numbers, we obtain a characteristic value of 
\begin{equation}
    n_{0} \sim 4 \times 10^{13} \mathrm{cm}^{-3} 
    \Bigg( \frac{\lambda_{{\rm Edd},w}}{f_{V}} \Bigg)
    \Bigg( \frac{f_{c}}{0.1} \Bigg)^{-1}
    \Bigg( \frac{v_{0}}{c_{s}} \Bigg)^{-1}
    \Bigg( \frac{T_{e}}{10^4 \mathrm{K}} \Bigg)^{-1/2}
    \Bigg( \frac{r_{0}}{10^{11} \mathrm{cm}} \Bigg)^{-2},
\end{equation}
\noindent where $\lambda_{{\rm Edd},w} = \dot{M}_{w}/\dot{M}_{\rm Edd}$ and $f_{V}$ is the volume filling factor of the wind. Thus we expect densities in this transonic region of the wind to be on the order of $\sim 10^{13}$ cm$^{-3}$. For reference, the density near the mid-plane of a standard Shakura-Sunyaev disc at $R \simeq 10^{11}$~cm is $\simeq 10^{15} \mathrm{cm}^{-3}$ \citep{frank_king_raine}. 

\begin{figure}
 \centering
\includegraphics[width=\linewidth]{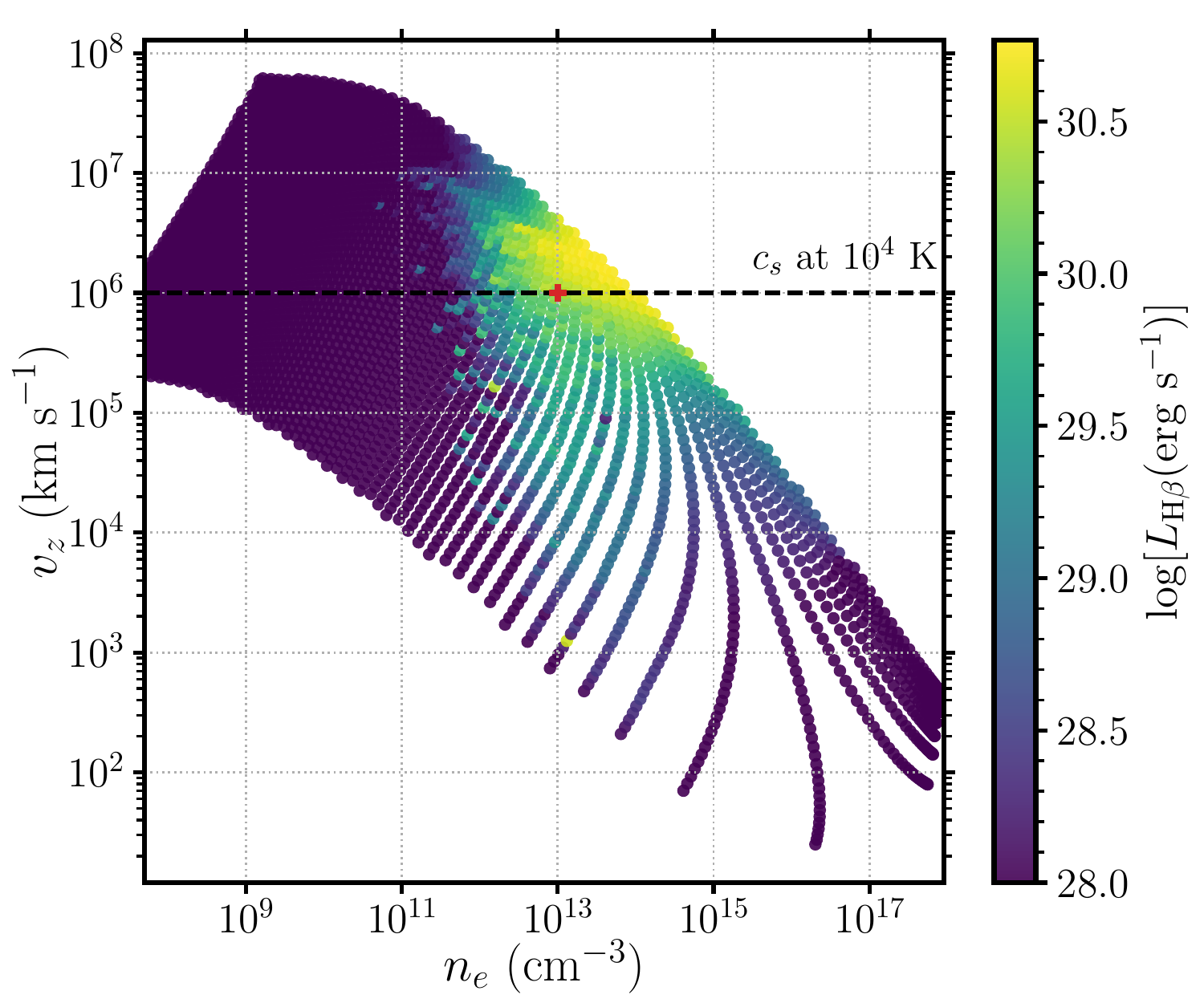}
 \caption{An illustration of the location of the line formation region in velocity-density parameter space for our fiducial model showing that the H$\beta$ line is formed in a transonic transition region between `atmosphere' and `wind'. For each point (corresponding to a wind grid cell), we plot the vertical velocity, $v_z$, against the electron density, $n_e$. The colour coding is set by the (logarithmic) local H$\beta$ luminosity; as described in the text, this luminosity does not necessarily escape to infinity but does escape the local Sobolev zone. The horizontal dashed line marks the sound speed at $10^4~{\rm K}$, and the red cross marks the estimates of `characteristic' conditions in the line formation region discussed in Section~\ref{sec:line_forming} and Section~\ref{sec:discuss_density}. 
 }
\label{fig:velocity_density}
\end{figure} 

We can now compare these numbers to the conditions found in the line-forming region of our fiducial wind model. These are illustrated in Figure~\ref{fig:velocity_density}, which shows the local H$\beta$ line luminosity as a function of both electron density, $n_e$, and vertical velocity, $v_z$. The location of $v = 10~{\rm km~s}^{-1}$~($c_s$ at $10^{4}~{\rm K}$), $n_e = 10^{13}~{\rm cm}^{-3}$ is marked with a red cross, and is roughly coincident with the region of parameter space from which most of the H$\beta$ line emission originates. These conditions are also in line with our median estimates in Section~\ref{sec:line_forming} and the expectations from the flat Balmer decrement (Section~\ref{sec:balmer_dec}). Thus the physical conditions our model requires for the optical line-forming region are broadly consistent with those that might be expected in the transition region between the hydrostatic disc atmosphere and a thermally-driven wind. To put it another way: our modelling implies that  emission from the base of a thermally-driven wind may self-consistently account for the recombination lines seen in the optical spectra of XRBs like \source. 

These results are promising, but it is also important to identify a few caveats. First, the association of the optical line-forming region with the base of a disc wind is not unique. Any model capable of generating similar physical conditions in the disc atmosphere will likely produce similar emission line spectra (e.g., from the temperature-inversion layer above an irradiated disc; \citealt{wu01}). After all, the optical line-forming region in our wind model lies well below the height at which material becomes unbound. Thus, the base of a hydrostatic, vertically extended atmosphere could account for the observed spectrum. Indeed, the conditions in this layer may be broadly agnostic to what is happening higher up in the atmosphere. This final point is important for our second caveat: our fiducial model is somewhat different in terms of its temperature structure and volume filling factor from the smooth thermal wind model that inspired it. Specifically, the wind never quite reaches the high (Compton) temperatures ($T_e \gtrsim 10^6~{\rm K}$) characteristic of the unbound, supersonic flow in thermal winds \citep{begelman83,done2018,higginbottom18}. However, the details of this outer wind region do not determine the emission from the dense transition region below it, and our modelling does not rule out a hotter supersonic flow. 

The above caveats are a useful reminder that what we {\em really} need at this point is a complete physical picture of the irradiated disc atmosphere, including its full radial and vertical structure. Developing such a picture will require self-consistent radiation-hydrodynamic simulations akin to those carried out by \cite{higginbottom18} and \cite{higginbottom20}, but including the physics needed to model the photosphere of the disc and covering the full range of relevant luminosities and spectral states. A particularly relevant study of X-ray-irradiated discs, \cite{waters21}, shows that under certain conditions, the disc-wind transition region is far from smooth and is instead fragmented and clumpy due to thermal instability. Such a clumpy, multi-phase transition region is precisely the line-forming region supported by our model. It may also be important for generally explaining the optical wind absorption features in XRBs. 

\subsection{Implications for other X-ray binaries}

Above, we have argued that the optical line emitting region may be a transition region between the disc atmosphere and the accretion disc wind. This is a promising identification since equatorial disc winds are thought to be associated with the soft X-ray states in XRB outbursts through the observations of highly ionized iron absorbers \citep{miller06,ponti12}. In addition, the observed conditions are in line with what has been seen in other XRBs. Flat Balmer decrements on the order of unity have been found earlier in the soft state of GX 339--4 \citep{rahoui14} and Aql X--1 \citep{panizoespinar21}, suggesting high densities and the line widths of the emission lines are compatible with velocities of Keplerian orbits in the outer disc.      

Based on our simulations, the flow has to be rather clumpy for the lines to arise from the transition region. This assumption is not unreasonable from a theoretical \citep[][see discussion in Section \ref{sec:discuss_density}]{waters21} or observational point-of-view. Considering the latter, in the high-inclination system Swift J1357.2$-$0933m fast variability optical dips were accompanied by blueshifted absorption troughs that can be attributed to a clumpy wind \citep{jimenezibarra19,charles19}. Similar absorption troughs, although much narrower, have been observed in \source\/ and MAXI J1348$-$630 \citep{panizoespinar22}. In \source, wind features in the form of absorption troughs and extended emission line wings were found during the soft state in the infrared lines Pa$\gamma$ and Pa$\beta$ by \citet{sanchezsierras20}. They note that the increased level of ionisation in the wind can cause the Balmer absorption profiles to disappear, while the Paschen series would be less affected. This scenario requires some mechanism to shield the strong X-ray emission from completely over-ionizing the wind. One possibility is having a clumpy and dense equatorial wind launched in the outer disc to retain suitable conditions for line formation.       

\begin{figure}
 \centering
\includegraphics[width=\linewidth]{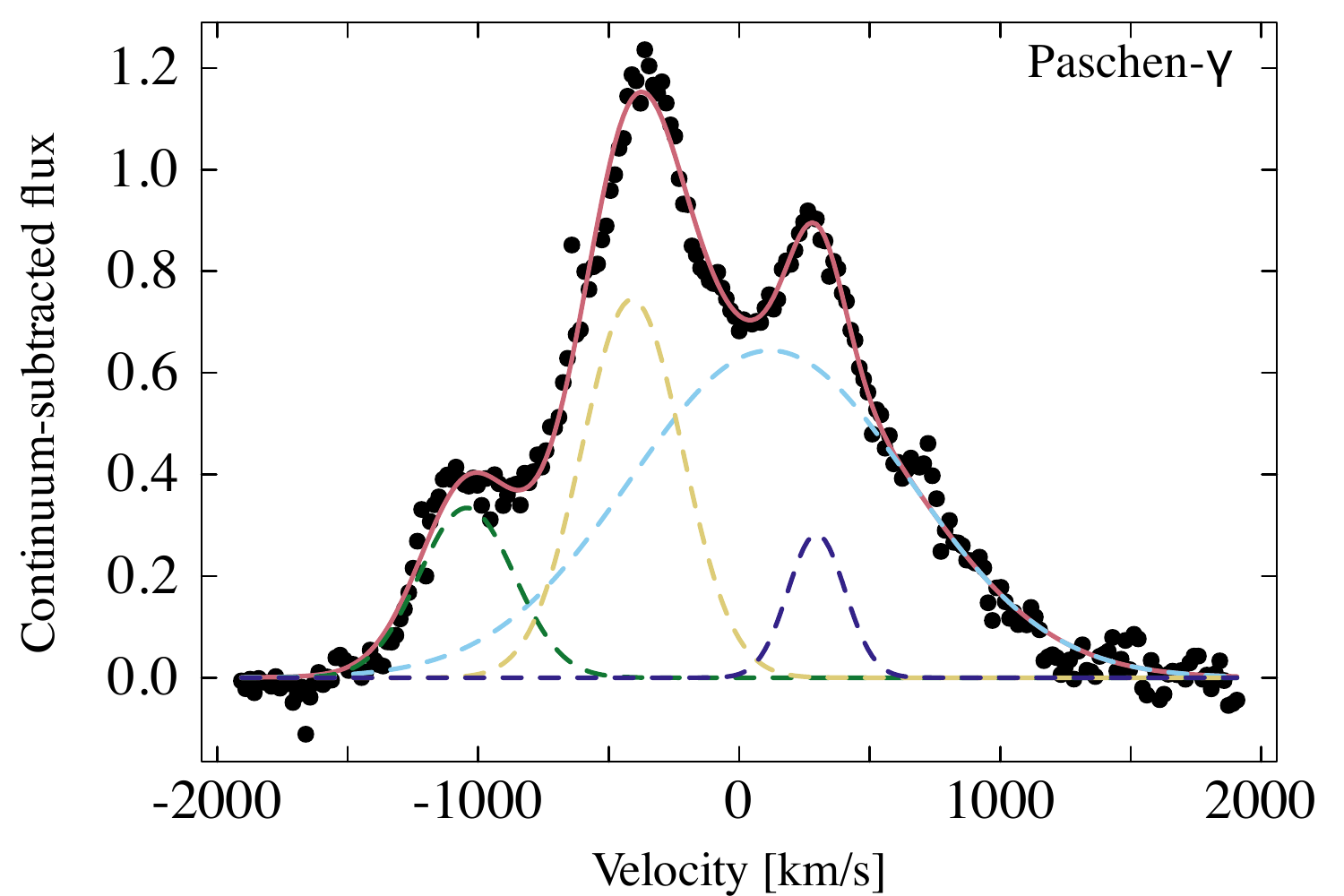}
 \caption{Continuum-subtracted line profile of the Paschen-$\gamma$ line at 10938 \AA. The profile can be fitted with four gaussian components indicating a complex line-emitting region.}
\label{fig:line_prof}
\end{figure}

As discussed in Section \ref{sec:sim_results_line},  the line profiles produced by our simulation qualitatively resemble the line profiles in the X-Shooter spectrum of \source, but they do not produce them in detail. In part, this is because some of the profiles have multiple peaks. In Fig. \ref{fig:line_prof}, we plot the line profile of Pa$\gamma$ from our X-Shooter observation. The profile is complex and can be fitted with four Gaussian components with varying velocities. Clearly, an azimuthally-symmetric rotating region cannot be responsible for all components. Other XRBs with wind features in their spectra also tend to present excesses in the blue/red tails of the emission lines relative to a single Gaussian fit. In many cases, these wings are non-symmetric, with the blueshifted side dominating. This has been observed in J1803$-$298 \citep{matasanchez22}, GX 339--4 \citep{rahoui14}, and MAXI J1348$-$630 \citep{panizoespinar22}.      

The observations of P-Cygni-like absorption profiles in the optical or near-infrared lines in systems with orbital inclinations surpassing 60 degrees in the hard states of XRBs \citep{bandyopadhyay97,munozdarias16,munozdarias17,munozdarias18,munozdarias19} indicate that the disc wind might be a more permanent structure in the evolution of XRB outbursts. What is changing is the amount of X-ray heating the outer disk due to changes in the geometry of the inner accretion flow between hard and soft states. This likely changes the temperature profile and the vertical structure of the outer disk with changing location and physical properties of the line-emitting region.     

\section{Summary} \label{conclusions}

We have analysed the optical/IR spectrum of the transient X-ray source \source\/ as observed with X-Shooter on 2018-07-13 when the X-ray source was in the soft state and dominated by emission from the disc. In order to carry out the analysis, we first presented the X-Shooter spectrum and used quasi-simultaneous \swiftuvot, \nustar, and \nicer\/ data to construct the SED for the source.  Then, based on relatively well-determined values of the mass, distance, reddening, and inclination, we developed a model of an X-ray illuminated disc that reproduces the SED of \source\/ at the time of the X-Shooter observations. Finally, we used the implied temperature distribution as a function of radius as an input to simulate the optical/IR spectrum in terms of a biconical outflow from the surface of the disc. These simulations were performed with a Monte Carlo radiative transfer code capable of determining the ionisation state and emergent spectrum of such systems.  By adjusting the parameters of the model, we were able to generate simulated spectra that qualitatively resemble the observations (Fig.~\ref{fig:python_spectrum}).

Our main conclusions are as follows:
\begin{itemize}
\item From a purely observational perspective, the optical/IR X-Shooter spectrum (Fig.~\ref{xs})  contains a rich collection of prominent, primarily double-peaked emission lines associated with Balmer, Paschen, and Brackett series of hydrogen as well as various lines of singly- and doubly-ionized helium. Bowen fluorescence lines and the \ion{Ca}{ii} H\&K lines are also readily apparent.  However, we do not find prominent P-Cygni-like profiles that would be expected from a wind moving at velocities significantly greater than that expected from disc rotation.

\item The multiwavelength SED of \source\ in the soft state can be modelled with an irradiated disc, soft X-ray thermal component, and a broken power-law to describe the UVOIR (Fig.~\ref{bbcont}). The latter components may arise from reprocessing in the wind or disc atmosphere. We did not find a suitable fit using only an irradiated disk component which underpredicted the near-infrared spectrum.    
\item The optical/IR emission lines observed in \source\/ arise in a dense region toward the outer disc (Fig.~\ref{fig:lines}). In our modelling, the lines are formed in the dense base of a biconical disc wind, in a region characterised by $T_e\sim40,000~{\rm K}$, $n_e\sim 10^{13}~{\rm cm}^{-3}$, and outflow velocities close to the sound speed (Fig.~\ref{fig:velocity_density}). The emission line region has to be dense, independent of detailed modelling, because the ratio of the Balmer lines requires high density (Fig.~\ref{fig:balmer_decrement}). Any {\em outflow} (as opposed to rotational) velocities must be small in the line-forming region. 

\item Much/most of the observed optical/IR continuum arises from the same region that produces the emission lines.  At the densities required to produce the emission lines, free-free and free-bound emission comprise the bulk of the continuum, especially at longer wavelengths.  Reprocessing by a wind or disc atmosphere can therefore have a profound effect on the optical (and even UV) continuum (Fig.~\ref{fig:python_sed}) and can explain the near-infrared excess in our observation (although even with the reprocessing component the data is still somewhat underpredicted in the near-infrared). When emission lines of this strength are observed in other XRBs, it is important to be aware of this effect when deriving the temperature profile of the disc.

\item Our modeling of the optical/IR emitting region in \source\/ cannot distinguish whether it arises from an atmosphere or an accretion disk wind. Our parameterisation for the wind does contain material moving at high velocity, but all of this material is too ionized and too tenuous to contribute to the optical/IR spectrum. Our modelling, combined with general estimates (section~\ref{sec:discuss_density}), is still consistent with the optical line and continuum emission originating from the dense base of a thermal wind or perhaps from a clumpy, unstable interface between the disc and wind. However, the optical/IR spectrum does not require the existence of fast, outflowing material. 
\end{itemize}
Overall, our work demonstrates that the transition region between the disc atmosphere and a thermally-driven wind can be the source of the rich optical line spectra seen in the soft states of X-ray binaries during outburst events. These \textit{wind bases} can also dramatically impact the infrared-to-UV continua, even dominating the emission at the red part of the spectrum. With the ubiquity of accretion disc winds and strongly irradiated discs in many astrophysical sources, we suggest that these effects also apply to other X-ray binaries in addition to \source, as well as other classes of accreting systems.         

\section*{Acknowledgements}
We thank Noel Castro Segura, Maria Georganti, Teo Mu{\~n}oz-Darias and Nathalie Degenaar for valuable discussions on optical winds in XRBs. We gratefully acknowledge Chris Done for providing us with the disc irradiation model \textsc{optxrplir} and Noel Castro Segura for further discussion of the model. This project has received funding from the European Research Council (ERC) under the European Union’s Horizon 2020 research and innovation programme (grant agreement No. 101002352). Partial support for KSL's effort on the project was provided by NASA through grant numbers HST-GO-15984 and HST-GO-16066 from the Space Telescope Science Institute, which is operated by AURA, Inc., under NASA contract NAS 5-26555. JM acknowledges funding from the Royal Society. This research is based on observations collected at the European Southern Observatory under ESO programme 0101.D-0356. This research has made use of data and/or software provided by the High Energy Astrophysics Science Archive Research Center (HEASARC), which is a service of the Astrophysics Science Division at NASA/GSFC.
We gratefully acknowledge the use of the following software packages: astropy \citep{astropy2013,astropy2018}, matplotlib \citep{matplotlib}, scipy \citep{2020SciPy-NMeth} and OpenMPI \citep{openmpi}.

\section*{Data Availability}
 The VLT/X-Shooter data analyzed here are available at the European Southern Observatory Science Archive Facility (http://archive.eso.org). UV and X-ray data are available at HEASARC. The simulation data produced as part of this work are available from the authors on request. 



\bibliographystyle{mnras}
\bibliography{references,software} 

\bsp	
\label{lastpage}
\end{document}